\documentclass[useAMS,usenatbib]{mn2e}
\usepackage{amssymb}
\usepackage{natbib}
\usepackage{times}
\usepackage{amsmath}
\usepackage{color}
\usepackage[pdftex]{graphicx}
\usepackage{amsmath}
\usepackage{amssymb}
\usepackage{comment}
\usepackage{supertabular}
\usepackage{hhline}
\usepackage{multirow}
\usepackage{booktabs}

\newcommand{\hii}{H{\sc i} 21\,cm}

\title {The Expanded Giant Metrewave Radio Telescope}
\author [Patra et al.]{N. N. Patra$^{1}$, N. Kanekar$^{1}$\thanks{Swarnajayanti Fellow; nkanekar@ncra.tifr.res.in}, 
J. N. Chengalur$^{1}$, R. Sharma$^{1}$, M. de Villiers$^{2}$, B. Ajit 
\newauthor 
Kumar$^1$, B. Bhattacharyya$^1$, V. Bhalerao$^1$, R. Bombale$^1$, K. D. Buch$^1$, B. Dixit$^1$, 
\newauthor 
A. Ghalame$^1$, Y. Gupta$^1$, P. Hande$^1$, S. Hande$^1$, K. Hariharan$^1$, R. Kale$^1$, S. Lokhande$^1$, 
\newauthor 
S. Phakatkar$^1$, A. Prajapati$^1$, S.~K.~Rai$^1$, P. Raybole$^1$, J. Roy$^1$, A. K. Shaikh$^1$, 
\newauthor
and S. Sureshkumar$^1$\\
$^{1}$ National Centre for Radio Astrophysics, Tata Institute of Fundamental Research, Pune University, Pune 411007, India \\
$^2$ South African Radio Astronomy Observatory, Black River Park, Observatory, 7925, South Africa
}
\date {}
\begin {document}
\maketitle

\begin{abstract}

With 30 antennas and a maximum baseline length of 25~km, the Giant Metrewave Radio Telescope 
(GMRT) is the premier low-frequency radio interferometer today. We have carried out a study of 
possible expansions of the GMRT, via adding new antennas and installing focal plane arrays (FPAs), 
to improve its point-source sensitivity, surface brightness sensitivity, angular resolution, field 
of view, and U-V coverage. We have carried out array configuration studies, aimed at minimizing the 
number of new GMRT antennas required to obtain a well-behaved synthesized beam over a wide range 
of angular resolutions for full-synthesis observations. This was done via two approaches, tomographic 
projection and random sampling, to identify the optimal locations for the new GMRT 
antennas. We report results for the optimal locations of the antennas of an expanded array 
(the ``EGMRT''), consisting of the existing 30 GMRT antennas, 30 new antennas at short distances, 
$\lesssim 2.5$~km from the GMRT array centre, and 26 additional antennas at relatively long 
distances, $\approx 5-25$~km from the array centre. The collecting area and the field of view of 
the proposed EGMRT array would be larger by factors of, respectively, $\approx 3$ and $\approx 30$, 
than those of the GMRT. Indeed, the EGMRT continuum sensitivity and survey speed with $550-850$ MHz 
FPAs installed on the 45 antennas within a distance of $\approx 2.5$ km of the array centre would be 
far better than those of any existing interferometer, and comparable to the sensitivity and survey 
speed of Phase-1 of the Square Kilometre Array.
\end{abstract}


\section{Introduction}
\label{intro}

At frequencies ranging from tens of MHz to hundreds of GHz, radio astronomy provides an outstanding 
and a unique window to study a wide range of astrophysical objects and phenomena. This includes pulsars, 
atomic, molecular and ionized gas in the Milky Way and other galaxies,
the environments of supermassive black holes and accretion disk physics, 
jets and lobes from active galactic nuclei (AGNs), protoplanetary disks, complex 
organic molecules, solar and planetary emission, galaxy clusters, the cosmic microwave background, the 
epoch of reionization, etc. Over the last five decades, radio interferometers such as the Westerbork 
Synthesis Radio Telescope (WSRT), the Very Large Array (VLA), the Australia Telescope Compact 
Array (ATCA), the Very Long Baseline Array (VLBA), etc., consisting of multiple dishes spread 
over distances much larger than the dish size, have used the technique of earth-rotation aperture 
synthesis \citep{mccready47,ryle52,obrien53} to obtain angular resolutions many orders of magnitude 
finer than would have been possible with even the largest single dish radio telescopes. Such 
telescopes have yielded detailed information on the structure and kinematics of Galactic and 
extra-galactic radio sources and have dramatically improved our understanding of the Universe.
More recently, over the last few years, there has been a dramatic upsurge in radio astronomy, 
with new interferometers such as the Atacama Large Millimeter/sub-millimeter Array \citep[ALMA; ][]{wootten09}
at high frequencies ($\gtrsim 100$~GHz), and the Low Frequency Array \citep[LOFAR; ][]{lofar13} and 
the Murchison Widefield Array \citep[MWA; ][]{mwa13} at very low frequencies ($\lesssim 300$~MHz). In 
parallel, there have been significant upgrades to the VLA, resulting in the Karl G. Jansky VLA (JVLA) which 
has outstanding sensitivity and frequency coverage at intermediate frequencies $\approx 1-50$~GHz 
\citep[e.g.][]{perley11}. New or upgraded interferometers with wide fields of view and survey speed, such 
as the Australian SKA Pathfinder \citep[ASKAP; e.g. ][]{johnston08}, the MeerKAT array \citep[e.g.][]{booth09}, 
and the APERTIF system on the WSRT \citep[e.g.][]{verheijen08} are being commissioned today. Finally, the 
next generation radio interferometer, the Square Kilometre Array (SKA), of unparalleled sensitivity, is 
currently being designed, aiming for completion in the next decade \citep[e.g.][]{dewdney15}.

\begin{figure}
\begin{center}
\begin{tabular}{c}
	\includegraphics[height=3.0in]{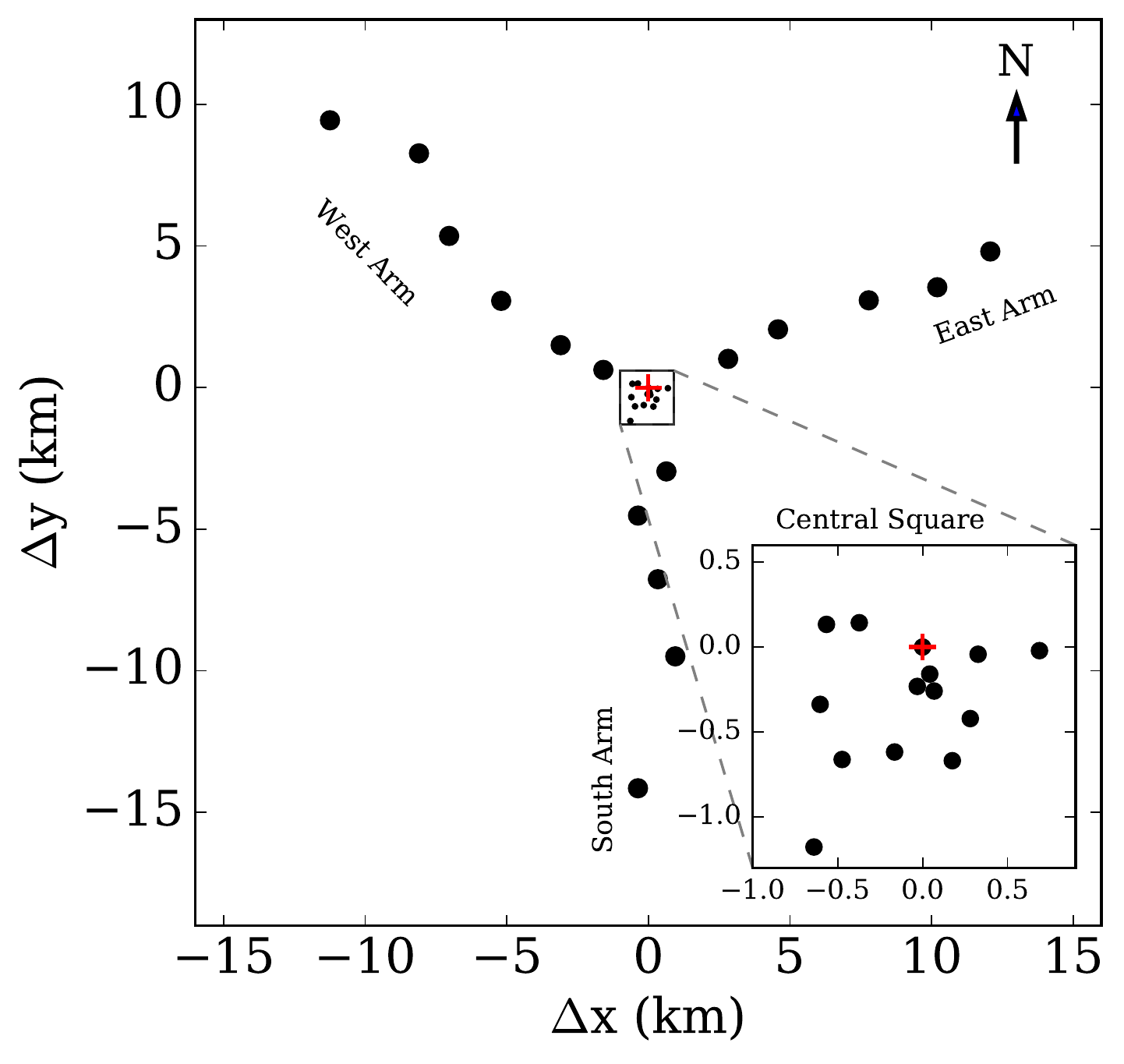}
\end{tabular}
\end{center}
\caption{The GMRT array, with antenna locations shown by the filled circles. The 30 GMRT antennas are 
arranged in a ``Y'' array, consisting of 14 antennas in a central core (the ``central square''), and 
the remaining 16 antennas in the three arms (``West'', ``East'', and ``South'') of the Y. The red ``+'' 
sign indicates the nominal origin of the array, at antenna C02.
\label{fig:gmrt}}
\end{figure}

The Giant Metrewave Radio Telescope \citep[GMRT; ][]{swarup91} is one of the leading radio 
interferometers in the world today. Completed around 20 years ago, the GMRT consists of 30 parabolic 
dishes, each of 45-m diameter, and with a longest baseline of $\approx 25$~km. Fourteen of the GMRT 
antennas are located within a ``central square'', of size $\approx 1 \; {\rm km} \; \times 1$~km, while 
the remaining 16 antennas are distributed along three arms of a Y, each of $\approx 12$~km length. This hybrid 
configuration, shown in Fig.~\ref{fig:gmrt}, with both closely-separated and distant antennas, yields good 
U-V coverage out to the longest antenna separations (i.e. $\approx 25$~km) over a full-synthesis track, and 
thus simultaneously provides information on radio emission at both small and large angular scales. 
The GMRT's observing frequencies lie in the range $\approx 150-1450$~MHz, with the upper range 
overlapping with the lower range of the VLA ($\approx 1-50$~GHz). The two arrays thus provide complementary 
views of the Universe, over a wide and contiguous range of frequencies. 

The GMRT has produced outstanding science in a wide range of areas, including pulsar studies 
\citep[e.g.][]{gangadhara01,freire04,hermsen13,bhattacharyya13,bhattacharyya16,roy15}, \hii\ spectroscopy 
of dwarf galaxies \citep[e.g.][]{begum06,begum08,roychowdhury10}, \hii\ and hydroxyl (OH) absorption studies of 
high-redshift galaxies \citep[e.g.][]{kanekar02,kanekar03,kanekar09b,kanekar14}, studies of 
AGNs and their environments \citep[e.g.][]{ishwar03,gupta06,lal07,aditya16}, studies of galaxy clusters 
\citep[e.g.][]{venturi07,venturi08,brunetti07,brunetti08,giacintucci11,vanweeren10,kale15,vanweeren17}, physical 
conditions in atomic gas in the Milky Way \citep[e.g.][]{kanekar11b,roy13b}, extra-galactic continuum studies 
\citep[e.g.][]{garn07,ibar09,ishwar10,mauch13,taylor16}, Galactic Plane studies 
\citep[e.g.][]{bhatnagar00,chengalur03b,roy04,roy13c}, studies of transient sources 
\citep[e.g.][]{vadawale03,chandra04,hyman09,roy10,chandra17}, giant radio sources 
\citep[e.g.][]{bagchi07,bagchi14,tamhane15,sebastian18}, \hii\ emission stacking studies of 
cosmologically-distant galaxies \citep[e.g.][]{lah07,lah09,kanekar16,rhee16}, 
constraints on fundamental constant evolution \citep[e.g.][]{chengalur03,kanekar10}, all-sky surveys 
\citep{intema17}, studies of the epoch of reionization \citep[e.g.][]{paciga11,paciga13}, etc.
At present, the GMRT is the premier telescope in the world in terms of sensitivity and angular resolution 
at low frequencies, $\lesssim 1$~GHz, and, indeed, has the largest collecting area of any fully
steerable telescope at all frequencies.

\section{The Upgraded GMRT}

All of the above studies were based on observations with the original GMRT, with a maximum bandwidth of 
$\approx 32$~MHz, and with narrow frequency bands, covering $\approx 150-156$~MHz, $\approx 230-245$~MHz, 
$\approx 300-360$~MHz, $\approx 570-660$~MHz, and $\approx 900-1450$~MHz. The GMRT is currently 
being upgraded, with the installation of new receivers covering $\approx 125-250$~MHz, 
$\approx 250-500$~MHz, $\approx 550-850$~MHz, and $\approx 950-1450$~MHz (i.e. near-seamless coverage 
over $\approx 125-1450$~MHz), and a new wideband correlator with a bandwidth of 400~MHz \citep{gupta17}. 
This will result in a significant increase in the telescope sensitivity (by a factor of $\approx 3$)
for continuum and pulsar studies, in the 
U-V coverage of the array for continuum studies of complex sources, and in the frequency coverage for 
studies of redshifted \hii\ and OH emission and absorption, and radio recombination lines.  Indeed, 
the installation of the first few GMRT $250-500$~MHz receivers resulted in two new detections of 
redshifted \hii\ absorption at $z \approx 2$ \citep[][]{kanekar14c}.

\begin{figure}
\begin{center}
\begin{tabular}{c}
	\includegraphics[height=3.0in]{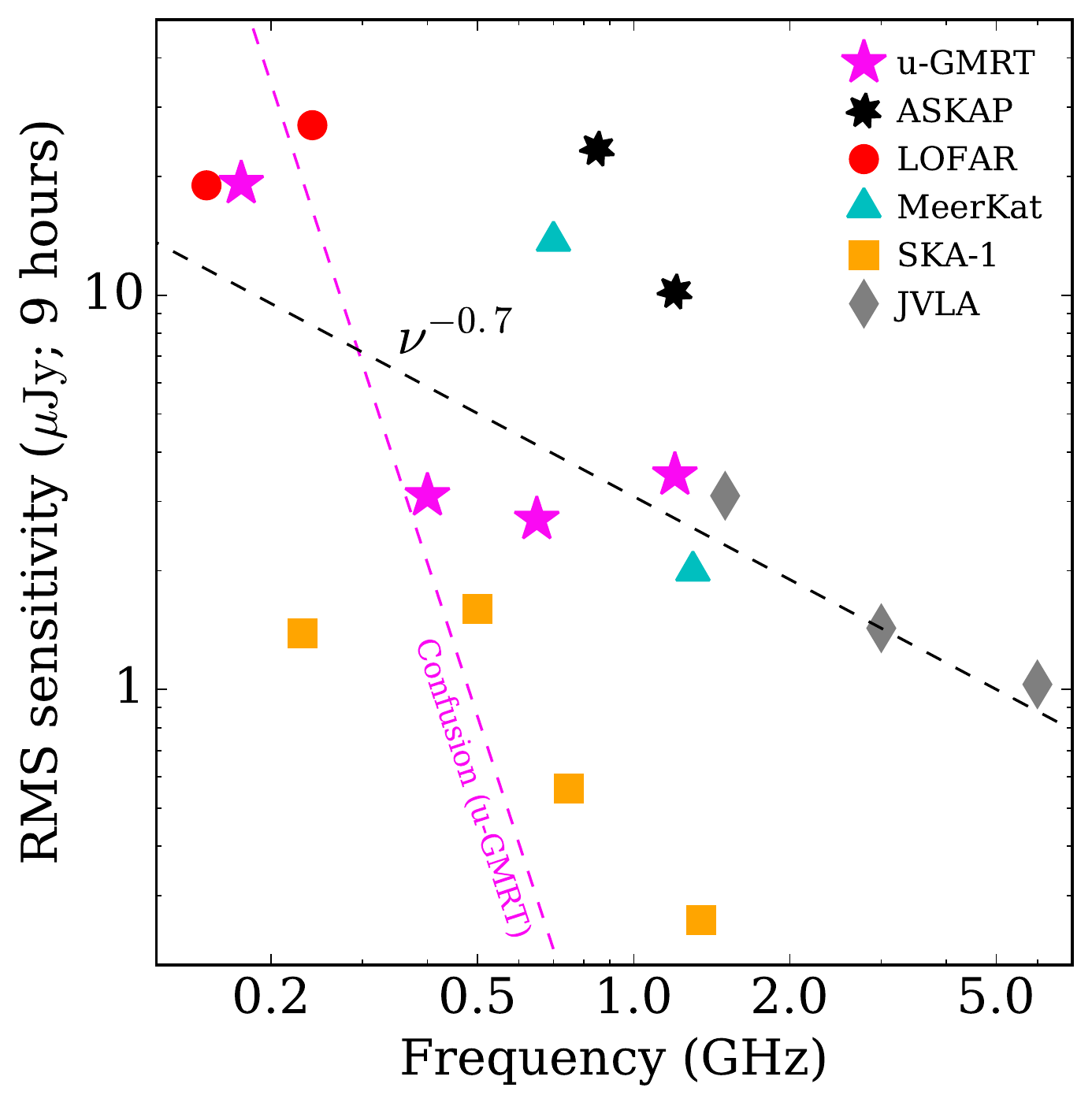}
\end{tabular}
\end{center}
\caption{The $1\sigma$ continuum noise of the uGMRT (in magenta stars) as a function of 
observing frequency, compared with the sensitivity of the best current and planned radio 
interferometers in the world [JVLA in grey stars, LOFAR in red circles, MeerKAT (in cyan 
triangles) , ASKAP (in black asterisks) and the SKA-1 (orange squares)], for a 9-hour full-synthesis 
integration. The dashed magenta line shows the $1\sigma$ GMRT confusion noise at the different 
observing frequencies. The dashed black line shows the spectral energy distribution 
of a typical synchrotron-spectrum extragalactic continuum source, with flux density 
$\propto \nu^{-0.7}$. The continuum sensitivities are from the webpages of the various telescopes
or, for existing arrays like the JVLA, from their Exposure Time Calculators; note
that the sensitivity of the MeerKAT array will be limited by source confusion, 
especially at frequencies $\lesssim 1$~GHz. See main text for discussion.}
\label{fig:ugmrt}
\end{figure}

Fig.~\ref{fig:ugmrt} compares the continuum sensitivity [i.e. the root-mean-square (RMS) noise] of the 
upgraded GMRT (the ``uGMRT'') for a 9-hour full-synthesis integration with the sensitivities of the best 
radio interferometers in the world, at frequencies $\lesssim 10$~GHz. The figure includes existing 
interferometers (the uGMRT, the JVLA, and LOFAR), interferometers that are now coming online (MeerKAT 
and ASKAP) and Phase~1 of the SKA (labelled ``SKA-1''). 
It also includes the $1\sigma$ source confusion limit (see Section~\ref{sec:long}) of the uGMRT at 
its different observing frequencies, shown as the magenta dashed line \citep[using equation~27 
of ][]{condon12}. The black dashed line shows the spectrum of a typical extra-galactic source emitting 
optically-thin synchrotron radiation, with a spectral index of $\alpha = -0.7$ (assuming that the source 
flux density $S_\nu \propto \nu^\alpha$). It is clear that the continuum sensitivity of the uGMRT over  
the frequency range $\approx 300 - 1400$~MHz will be sufficient to detect all typical synchrotron-emitting 
sources detected by the JVLA at frequencies $\gtrsim 1$~GHz; the two telescopes will thus continue to 
complement each other. However, it is also clear that the sensitivity of the uGMRT at frequencies 
below 500~MHz will be limited by source confusion in even relatively short integrations. Further, 
the MeerKAT array will have a sensitivity (limited by source confusion) comparable to that of 
the uGMRT (and the JVLA) at frequencies $\gtrsim 1$~GHz, while the SKA-1 would have a far better 
sensitivity than the uGMRT throughout its frequency range.

Another important metric characterizing a modern radio telescope is survey speed; the large 
diameter of the GMRT dishes imply that the uGMRT's survey speed figure of merit is lower than 
that of an interferometer like MeerKAT which has smaller dishes, and far lower than that of 
ASKAP, which has both smaller dishes and phased-array feeds \citep[see, e.g., Table~1 
of ][]{dewdney15}. Wide-field surveys would thus require a far larger number of uGMRT 
telescope pointings  and thus a concomitant increase in observing time.

Finally, the increased GMRT bandwidth has meant a significant improvement in sensitivity 
for both continuum and pulsar studies. However, while the improved frequency coverage implies 
a significant increase in the redshift range accessible for GMRT studies in the redshifted 
\hii\ and OH lines, there has been no increase in the sensitivity for such spectral line studies.
This would only be possible by increasing the number of antennas and/or by decreasing the
system temperature.

In summary, while the uGMRT will definitely yield outstanding science over the next decade, 
it is important to begin considering the next expansion of the telescope, to retain its 
importance in the SKA era. In this paper, we discuss different strategies for an expanded GMRT
(the ``EGMRT'') and their science benefits, and finally describe the results of array 
configuration studies for the locations of the new antennas of the expanded array.

\section{The Expanded GMRT}
\label{sec:egmrt_main}

We assume that the frequency coverage of the GMRT will remain approximately unchanged 
in the expansion (except possibly for a minor extension to lower frequencies, $< 100$~MHz). This is 
because the mesh spacing of the current GMRT antennas would imply a rapid drop in sensitivity 
at frequencies $\gtrsim 2$~GHz. Note that a reduction in the mesh spacing of the existing GMRT antennas 
would increase the wind loading, and hence could affect the structural stability of the dishes.
We first consider the basic issue of the point-source sensitivity of the present GMRT and compare 
this with the sensitivities of current and planned arrays. We then considered three possible 
avenues for the expansion of the GMRT, to achieve (1)~a wider field of view, (2)~improved surface 
brightness sensitivity, and (3)~improved angular resolution, and hence a better confusion limit. 
The broad science drivers and possible challenges for each of these approaches are discussed in 
brief in the present section.

\subsection{Point Source Sensitivity}
\label{sec:sensitivity}

The original GMRT was built with the aim of being sufficiently sensitive to detect \hii\ line emission 
from neutral hydrogen from massive proto-clusters at $z \approx 3$, a prediction of the hot dark matter 
cosmological model \citep{swarup91}. The continuum sensitivity was sufficient to provide a low-frequency 
counterpart to the original higher-frequency VLA, allowing the detection at frequencies $\lesssim 1.4$~GHz 
of synchrotron emission from extragalactic sources detectable with the VLA at frequencies $\gtrsim 1.4$~GHz. 
It can be seen 
from Fig.~\ref{fig:ugmrt} that the uGMRT will provide a similar low-frequency counterpart to the JVLA, 
with comparable continuum sensitivity at $\approx 1.4$~GHz. However, it is also clear from the figure that 
the uGMRT will have a far lower sensitivity than SKA-1 at all frequencies. While SKA-1 is likely to be 
built in the southern hemisphere, leaving the northern hemisphere niche for the uGMRT, increasing the 
point-source sensitivity of the GMRT is critical for it to remain competitive over the next decade, 
in the SKA era. Fig.~\ref{fig:ugmrt} shows that this would require roughly a tripling of the uGMRT 
sensitivity.

\subsection{Focal plane arrays: An increased field of view}
\label{sec:fpa}

The GMRT currently has single-pixel feeds, and hence, a relatively small field of view, implying a
low survey speed, even with the wider bandwidths of the uGMRT. In recent years, much emphasis has 
been placed on the development of focal plane arrays (FPAs) with relatively low system temperatures 
\citep[e.g.][]{vanardenne09,deboer09,chippendale16}, to obtain both 
a wide field of view and a high point-source sensitivity. L-band FPAs covering $\approx 1000-1700$~MHz 
have been installed on the WSRT \citep[the APERTIF system; e.g.][]{verheijen08} and the new ASKAP 
array \citep[e.g.][]{johnston08}, each with 30 independent beams on the sky, resulting in far higher 
survey speeds than at the GMRT \citep[see Table~1 of][]{dewdney15}. Such an FPA system, with 
$\approx 30$-beams, would imply a huge jump in the GMRT's field of view. We note that the GMRT's prime-focus 
feeds are uncooled, with relatively high system temperatures, $T_{\rm sys} \approx 70-100$~K 
at $300-1450$~MHz. The high point-source sensitivity of the GMRT arises due to its large collecting 
area; the installation of FPAs on the GMRT would hence not give much of a penalty in system 
temperature.

The main science drivers for an FPA system on the GMRT are \hii, pulsar, and continuum surveys, 
and the exciting new field of radio transients, especially fast radio bursts \citep[FRBs; e.g.][]{lorimer07,thornton13,spitler16}. 
While a low-frequency FPA (at $\lesssim 500$~MHz) would significantly increase the GMRT survey speed, benefitting
radio continuum and pulsar surveys, it is very unlikely to be possible to detect redshifted \hii\ emission 
from individual galaxies at $z \gtrsim 1.8$, for which the \hii\ line would redshift to frequencies 
$\lesssim 500$~MHz. Further, while initial studies were unable to detect FRBs at frequencies 
$\lesssim 800$~MHz \citep[e.g.][]{petroff16}, FRBs have recently been detected at frequencies 
$\gtrsim 400$~MHz with the Canadian Hydrogen Intensity Mapping Experiment \citep{chime18}. The two broad 
frequency ranges of interest for an FPA system 
on the GMRT are hence likely to be $\approx 1000-1400$~MHz and $\approx 550-850$~MHz. FPAs covering the 
former frequency range have already been installed on the WSRT and ASKAP arrays; the best science outcomes
for the GMRT are hence likely to be obtained from an FPA covering the relatively unexplored frequency range 
of $\approx 550-850$~MHz. The large field of view and frequency coverage of such a system, coupled with 
GMRT's high sensitivity, would yield a large instantaneous survey volume, allowing one to detect \hii\ 
emission from individual massive galaxies 
at $z \approx 0.6-1.5$. Optical studies of star-forming galaxies have shown that the cosmic star formation rate (SFR) density
of the Universe peaks in the redshift range $\approx 1-3$, often referred to as the ``epoch of galaxy assembly'',
before declining by an order of magnitude to the present epoch \citep[e.g.][]{hopkins06,bouwens14}. 
However, little is known about atomic gas in these galaxies, the fuel for such star formation. The 
GMRT single-pixel feeds have been used to obtain an upper limit on the average gas mass of star-forming
galaxies at $z \approx 1.3$, by co-adding their \hii\ emission signals \citep{kanekar16}, but the 
GMRT field of view is too small to carry out a survey for \hii\ emission from massive galaxies at these 
redshifts in reasonable observing time. The prospect of using FPAs covering $550-850$~MHz on the GMRT to 
trace the evolution of atomic gas in star-forming galaxies from the peak epoch of star formation down 
to intermediate redshifts is hence a very exciting one (Patra et al., in prep.). Of course, a 30-beam FPA 
system at these frequencies would yield a field of view of $\approx 15$~square degrees at $\approx 600$~MHz, 
allowing wide-field high-sensitivity surveys for pulsars, FRBs, and extra-galactic continuum sources.

The main challenges in installing FPAs on the GMRT are the large data volumes, signal transport (especially 
bringing a large number of signal-carrying cables from the individual FPA elements to the base of each 
antenna), digital signal processing, the possibility of RFI on short baselines, and supporting an FPA 
at the prime focus of the GMRT antennas. None of these appear intractable at this time. We note, in passing,
 that, since the FPAs are likely to be installed mostly on the shorter-baseline antennas (see below), 
land acquisition is unlikely to be a critical issue for this expansion route.

\subsection{Short baselines: Surface brightness sensitivity}
\label{sec:short}

The GMRT currently has few antennas at distances $\lesssim 0.2$~km from each other, and hence has 
relatively poor U-V coverage at short U-V spacings, which adversely affects its ability to image
extended, large-scale radio emission. Indeed, we note that the GMRT has only three baselines 
with a physical separation $< 100$~m. This is a serious limitation for studies of complex fields in 
the Galactic plane, and especially of the exciting region around the Galactic Centre 
\citep[e.g.][]{anantha91, larosa00,nord04,yusufzadeh04}. Radio relics and halos in galaxy clusters 
are also typically very extended, requiring good U-V coverage at short spacings to both detect the 
emission and study its physical properties 
\citep[e.g.][]{venturi07,brunetti08,deo17}. Detecting ``cosmological halos'' of ionized gas around 
massive high-$z$ quasars would also require a large collecting area at short baselines 
\citep[e.g.][]{sholomitskii90,geller00}. Adding new antennas with short physical separations 
($\ll 1$~km) to the GMRT would significantly improve its surface brightness sensitivity.
Of course, both \hii\ emission and pulsar surveys would benefit from adding antennas at such short 
spacings. In the case of \hii\ emission studies of external galaxies, new antennas at short spacings
would not resolve out the emission. Conversely, for pulsar surveys, the number of phased-array beams 
needed to cover the full primary beam would be very large if one were to include long-baseline antennas 
\citep[e.g.][]{roy18}; further, antennas on short baselines would be easier to phase up for pulsar 
searches.

The primary challenge for the short-baseline expansion is likely to be RFI, which would not decorrelate 
on short baselines. We note that RFI has been a steadily worsening problem at the GMRT over the last 
decade and that online RFI mitigation techniques \citep[e.g.][]{buch16} will be critical to deal 
with this issue. Since the observatory already owns most of the required land for the short-baseline 
expansion (for baselines $\lesssim 1.7$~km; see Section~\ref{sec:inputs}), land acquisition is unlikely 
to be a serious problem here.

\subsection{Long baselines: The confusion limit}
\label{sec:long}

The maximum attainable angular resolution of an aperture synthesis radio telescope is decided by its 
longest baseline. The GMRT currently has a longest baseline of $\approx$ 25~km, which implies an angular 
resolution of $\approx 3''/[\nu/{\rm GHz}]$ . This angular resolution sets the ``confusion limit'' of 
the array, the RMS noise arising due to the blending of multiple faint individually-undetected sources 
within the array synthesized beam \citep[e.g.][]{mills57,mitchell85,condon12}. It has long been appreciated 
that source confusion plays an important role in determining the continuum sensitivity of a synthesis 
telescope \citep[e.g.][]{mills57,scheuer57}; for deep continuum images, the detection threshold is set by a 
combination of the theoretical RMS noise and the confusion noise. 

For fifty years, arrays have mostly been designed so as to not be limited by source confusion. However, 
the huge increase in the bandwidth of existing radio telescopes, due to advances in signal transport 
methods and correlator capacity, without a corresponding increase in the baseline length 
has meant that the continuum sensitivity of today's interferometers 
is often limited by source confusion at low frequencies ($\lesssim 1$~GHz), rather than by the theoretical 
RMS noise. The best estimate of the low-frequency confusion limit can be obtained by extrapolating equation~(27) 
of \citet{condon12} 
\begin{equation}
\rm \sigma_c^* \approx 1.2 \mu Jy Beam^{-1} \left[ \frac{\nu}{3.02 \; GHz}\right]^{-0.7} \left[ \frac{\theta}{8''}\right]^{10/3} 
\end{equation}
to the observing frequency, $\nu$, where $\theta$ is the full-width-at-half-maximum (FWHM) of the array synthesized beam, and 
$5\sigma_c^*$ gives an estimate of the source detection threshold due to confusion. Note that $\sigma_c^*$ is 
not the rms confusion noise \citep[see the discussion in][for the definition of $\sigma_c^*$]{condon12}; 
indeed the distribution is highly skewed so the RMS confusion noise does not give a good 
estimate of the detection threshold \citep[][]{condon12}. For the GMRT, this implies $\rm \sigma_c^* \approx 0.052 
\times (\nu/1\;{\rm GHz})^{-4.0}$, i.e. a $5\sigma$ detection threshold of $\approx 24 ~\mu$Jy at 327~MHz. 
This was not a serious issue for the original GMRT , with a bandwidth of 32~MHz, as the $5\sigma$ 
detection threshold in a full-synthesis 327~MHz run was $\approx 50 \; \mu$Jy, implying that full-synthesis 
327~MHz images were not significantly limited by source confusion.  
However, it is clear from Fig.~\ref{fig:ugmrt} that $\rm \sigma_c^*$ is comparable to the theoretical RMS 
noise at $\approx 400$~MHz for a full-synthesis run, implying that the uGMRT would be limited by 
source confusion in the $250-500$~MHz band for observing times $\approx 10$~hours. The only way to address
this issue is to increase the angular resolution of the array, by installing antennas at long baselines, 
$> 25$~km. The strong dependence of $\rm \sigma_c^*$ on the FWHM of the synthesized beam 
\citep[$\sigma_c^* \propto \theta^{10/3}$;][]{condon12}
implies that merely doubling the angular resolution reduces the confusion limit by an order of 
magnitude. Specifically,
increasing the length of the longest GMRT baseline to $\approx 50$~km would reduce $5\sigma_c^*$ at 400~MHz
to $\approx 2.3 ~\mu$Jy, below the theoretical RMS noise ($\approx 3 ~\mu$Jy) in a full-synthesis uGMRT 
observation at a central frequency of 400~MHz. Doubling the length of the longest GMRT baseline would thus 
render source confusion an issue only for extremely deep 400-MHz integrations ($\gg 100$~hours). We will 
hence use 50~km as the target length of the longest baseline of the expanded array.

Acquiring the land needed for the installation of the new antennas is likely to be the biggest 
challenge for the long-baseline expansion. We are currently carrying out a land survey for this 
purpose, to identify tracts of land that might be used as antenna sites. Signal transport from 
the distant antennas is unlikely to be a serious problem, while RFI, although always an issue, 
should have its weakest effects on the long baselines.

\subsection{The Proposed Expanded GMRT}
\label{sec:egmrt}

The original GMRT was also designed to be suitable for multiple science goals, and hence has roughly half its 
collecting area in the central regions and half on the long baselines, out to 25~km. The former is useful 
for \hii\ emission studies of nearby galaxies and pulsar studies, besides providing acceptable surface 
brightness sensitivity for Galactic plane studies, while the latter yields the angular resolution needed 
to overcome source confusion and produce deep images of extra-galactic fields, as well as spatially-resolved 
information on individual sources like radio galaxies or galaxy clusters. In the case of the expanded
GMRT, we aim to increase the point-source sensitivity by a factor of $\approx 3$, by adding new antennas 
to the array. Further, as discussed in Sections~\ref{sec:short} and \ref{sec:long}, there are 
excellent science arguments for adding the new antennas on both short and long baselines. Since 
none of these science drivers appears to dominate over the others, we plan to retain GMRT's 
multi-science capabilities, and distribute the new antennas on both short and long baselines, 
so as to significantly improve the surface brightness sensitivity, the sensitivity to high-$z$ 
\hii\ emission, the pulsar sensitivity, and the confusion limit. This may be accomplished by adding roughly 
half the new collecting area (i.e. $\approx 30$~antennas) in a central region of size $\approx 5$~km, aiming 
to use this to provide excellent sensitivity for high-$z$ \hii\ emission studies, pulsar studies, and 
Galactic plane studies, and the remaining $\approx 30$ new antennas on intermediate and long baselines, 
to improve the angular resolution by a factor of $\approx 2$, and thus improve the confusion limit by 
a factor of $\approx 10$. This approach would triple the EGMRT's point-source sensitivity relative to 
that of the uGMRT, while also significantly improving both its surface brightness sensitivity and confusion
limit.

Next, it appears clear that installing FPAs with $\approx 30$~beams will be critical to obtaining a 
high survey speed, especially given the fact that the large diameter of the GMRT antennas implies a 
relatively small field of view for single-pixel feeds. Installing a 30-beam FPA on the full array would 
imply serious problems in signal transport from the more distant antennas to the central correlator. 
However, none of the main science drivers for a $550-850$~MHz FPA system (pulsar surveys, \hii\ emission 
from galaxies at $z \approx 0.6-1.5$, wide-field continuum surveys, and searches for fast radio 
transients) require the FPAs to be installed on long-baseline antennas. Installing a 30-beam FPA system 
on the core antennas of the new array, out to maximum baselines of $\approx 5-10$~km, would alleviate 
the signal transport issue. The confusion limit for wide-field continuum surveys with this sub-array, 
assuming a maximum baseline of $\approx 10$~km and equation~(27) of \citet{condon12}, would give a 
$5\sigma$ detection threshold of $\rm 5\sigma_c^* \approx 23 ~\mu$Jy at 700~MHz.

\section{The EGMRT Antenna Configuration}
\label{sec:locations}

As discussed in the previous section, we would like to explore the possibility of increasing 
the sensitivity of the uGMRT by a factor of $\approx 3$, by adding $\approx 30$~antennas in a central 
region, of size $\lesssim 10$~km, and a further $\approx 30$~antennas out to baselines of $\approx 50$~km. 
In order to significantly improve the surface brightness sensitivity, one has to also increase the number 
of antennas on very short spacings $\ll 1$~km. Our next step is to identify the locations of the proposed 
new antennas. The critical requirement here is that the antenna configuration provides a good U-V coverage, with no holes 
in U-V space that might give rise to high sidelobes in the array point spread function and hence, artefacts when imaging
complex fields. Our aim is hence to identify an antenna configuration that yields a synthesized beam as close to an 
``ideal'' beam as possible. We will use a 2-dimensional (2-D) circular Gaussian synthesized beam as the ideal beam for 
all configurations. Since the U-V coverage is the 2-D Fourier transform of the synthesized beam, we aim to 
obtain a U-V distribution as close to a 2-D Gaussian as possible, with an appropriate choice of the FWHM of 
this 2-D Gaussian. We then rank antenna configurations based on the fractional RMS difference between the actual 2-D 
U-V distribution and the ideal 2-D Gaussian distribution, aiming to minimize this fractional RMS difference. This 
quantity, expressed as a percentage, will be referred to as the ``Residual RMS''; note that a lower Residual 
RMS implies a better agreement between the actual and the ideal U-V configuration, and hence a preferred antenna 
configuration. 

It is important to also emphasize that, unlike in the case of a new array such as MeerKAT, ASKAP, or the SKA, we will 
be adding antennas to an existing array. This complicates the minimization procedure as the locations of the 
existing GMRT antennas must be frozen in the minimization.

We have alluded above to the critical issue of land availability in the area around the GMRT, an important 
constraint for the optimization. The GMRT already owns some land around the existing central square, out to baselines 
of $\approx 1.7$~km, that might immediately be used for an expansion. It should be possible to select the 
preferred antenna sites here, based on the simulations below for the optimal U-V coverage. We are currently carrying 
out a survey to identify land that may be acquired for the purpose of expanding the GMRT; this 
survey is now complete out to a region of $\approx 5$~km diameter around the central square. 
Land in the above two categories, i.e. either already owned by the GMRT or owned by the government and that 
might be acquired for the array expansion, was included in the allowed antenna locations in the simulations out to baselines of $\approx 5$~km. 
However, we emphasize that changes may be needed in the antenna configuration identified below, in case it is not 
possible to acquire individual locations. Finally, for baselines longer than $5$~km, we have chosen to identify the 
optimal antenna location via the present simulations. We note that, for the long-baseline antennas, the exact location 
of each individual antenna (within $\approx 1$~km) is unlikely to have a significant impact on the U-V coverage, and hence 
on the synthesized beam. We hence plan to carry out a land survey within $\approx 1$~km of each new optimal antenna location 
determined below to identify government land that might be acquired for the final antenna locations. We emphasize that 
the antenna locations identified by the approach below may not be the final ones, especially for the long-baseline 
antennas; however, the inferred antenna configurations will serve as a benchmark to test possible array configurations
based on the final land surveys, and additional constraints arising from optical fibre connectivity, accessibility, 
etc.

\subsection{The two approaches: Random sampling and Tomographic projection}
\label{sec:algorithms}

Many approaches exist in the literature to the problem of optimizing antenna locations for radio 
synthesis arrays \citep[e.g.][]{keto97,boone01,devilliers07}. We have chosen to use two separate, and independent, 
schemes to identify the array configuration that minimizes the Residual RMS. The first, applicable to relatively 
small areas (e.g. to the short- and intermediate-baseline configurations), is based on a simple Monte Carlo 
approach, in which we set up a grid of allowed antenna locations and then determine the Residual RMS for 
a large number (typically, $10^4$) of array configurations using random sampling of the possible locations, including 
any constraints based on land availability. The array configuration that yields the minimum Residual RMS is selected 
as the best configuration; we will refer to this as ``random sampling''. This method is computationally expensive, 
but is guaranteed to yield the best configuration, given a sufficiently large number of random samples, and 
hence works well for small areas where it is possible to sample a large fraction of the possible array configurations.

The second approach, referred to as ``tomographic projection'', is based on reducing the problem of two-dimensional
U-V coverage to a one-dimensional problem, by taking random projections of the two-dimensional antenna distribution 
along different angles and then moving the antennas so as to minimize the difference between the U-V distribution in 
one dimension (for each projection) and an ideal distribution \citep{devilliers07}. The antenna locations are shifted
for each projection direction, until one obtains an acceptable two-dimensional U-V coverage, and hence, an acceptable 
synthesized beam. Note that the minimization procedure is complicated by the fact that one would like the U-V points 
(i.e. the baseline distribution) to have an ideal distribution, but moving an antenna to shift a single U-V point 
also shifts all the other U-V points arising from that antenna \citep[i.e. the U-V points are not independent;][]{devilliers07}. 

The tomographic projection algorithm has been implemented in the {\sc iAntConfig} software package; we used this to optimize 
our antenna layouts. However, we found that the results of the {\sc iAntConfig} optimization are sensitive to 
the initial conditions, and that more stable results are obtained by combining the {\sc iAntConfig} minimization
procedure with a Monte Carlo approach, running {\sc iAntConfig} $\approx 100$ times for different initial antenna 
configurations. We evaluated the Residual RMS for each of the {\sc iAntConfig} output layouts, each with the difference 
between the actual and ideal U-V distributions minimized using tomographic projection, and chose the 
antenna configuration with the lowest value of the Residual RMS. We tested the results of this approach against those 
from the random sampling procedure for small regions (where the random sampling procedure is reliable), and found that
the two approaches yielded very similar array configurations.

We note, in passing, that the tomographic projection approach is significantly less computationally intensive than the 
random sampling method, and thus appears far better suited for optimizing array configurations. However, it is not 
straightforward to include constraints on land availability in the tomographic projection optimization. The current 
implementation in {\sc iAntConfig} carries out the optimization without including land constraints, and applies the 
land constraint at the end, by shifting the antenna locations to the nearest available ones. For sparse land availability 
(as is the case around the GMRT), this does not guarantee a minimum Residual RMS. The random sampling approach is 
not adversely affected by land constraints and, in fact, works better for sparse land availability because the number 
of allowed antenna locations is significantly reduced, making it possible to sample a larger fraction of possible
array configurations. We hence chose to use the random sampling approach for the short- and intermediate-baseline
antenna locations, but to use tomographic projection for the long-baseline locations.

\subsection{Inputs for the optimization}
\label{sec:inputs}

The critical inputs for the optimization are the observing frequency, the bandwidth, the number of channels, the 
time resolution, the total integration time, and the target declination. We chose to carry out the optimization at 
$\approx 1.2$~GHz in GMRT's highest observing frequency band, since the fractional bandwidth here is the worst for 
a given observing bandwidth. This implies the worst U-V coverage of all GMRT bands, and hence emphasizes any holes 
in the U-V coverage. 

We carried out the optimization for different time resolutions, to examine the effects of the selected 
sampling time on the derived array configurations. Of course, high temporal resolution would require 
significantly more computational time. No significant difference in the final array configuration was 
obtained on using resolutions finer than $\approx 120$~seconds. We hence finally used time resolutions 
of $\approx 120-180$~seconds for all optimizations, with the coarsest time resolution used for 
the shortest baseline configurations, and $120$~seconds used for all other optimizations.


Next, a large fractional bandwidth significantly improves the U-V coverage of a radio interferometer. 
The fractional bandwidth of the uGMRT ranges from $\approx 0.3-0.5$ at the different frequency 
bands, with the best fractional bandwidths at the $250-500$~MHz and $125-250$~MHz bands 
\citep{gupta17}. Again, carrying out the full array optimization with high frequency resolution 
is computationally very expensive. We hence chose to ignore the effects of a large fractional 
bandwidth in the optimization, and instead carried out the optimization for a single frequency 
channel. We emphasize that this is an extremely conservative approach, and that the ``true'' U-V 
coverage for the full EGMRT band would be significantly better than our single-channel estimate. 

The varying U-V coverage with target declination was handled by carrying out each optimization independently 
at four different declinations, $\delta = -30^\circ$, $0^\circ$, $+30^\circ$, and $+60^\circ$. The array configuration
obtained from each optimization was then applied to a wide range of declinations, from $-30^\circ$ to $+60^\circ$, 
and the final array configuration was chosen so as to yield the best average performance (i.e. the lowest 
Residual RMS) across the different declinations.

Finally, the GMRT antennas have an elevation limit of $17.5^\circ$, implying that most sources in the northern hemisphere 
are observable for $\gtrsim 10$~hours, while southern sources are observable for shorter periods, $\approx 7.5$~hours at 
$\delta = -30^\circ$. The array optimizations were carried out assuming a full-synthesis run at all declinations, i.e. $7.5$h
of total time at $\delta = -30^\circ$, $10{\rm h}45{\rm m}$ at $\delta = +30^\circ$, and $11{\rm h}45{\rm m}$ at 
$\delta = +60^\circ$. We also assumed, based on the settings for typical GMRT observations, that $\approx 85$\% of 
the time of a full-synthesis run is spent on the target source, and $\approx 15$\% on calibration.

\subsection{The Optimization: Strategy and Results}
\label{sec:strategy}

\begin{figure*}
\begin{center}
	\begin{tabular}{ccc}
\includegraphics[scale=0.375,trim={0.25cm 0.0cm 0.0cm 0.0cm},clip]{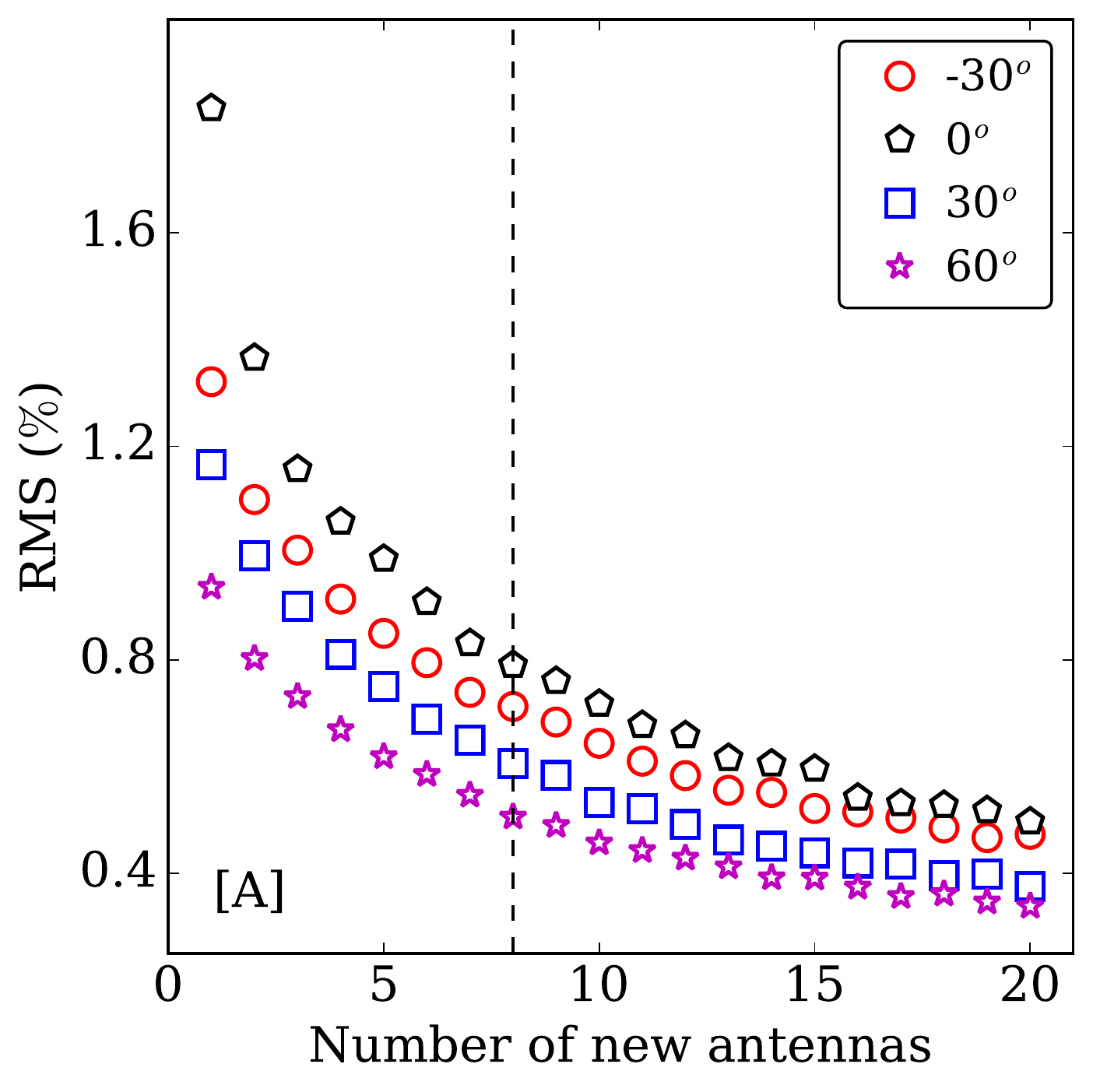} & 
\includegraphics[scale=0.375,trim={0.25cm 0.0cm 0.0cm 0.0cm},clip]{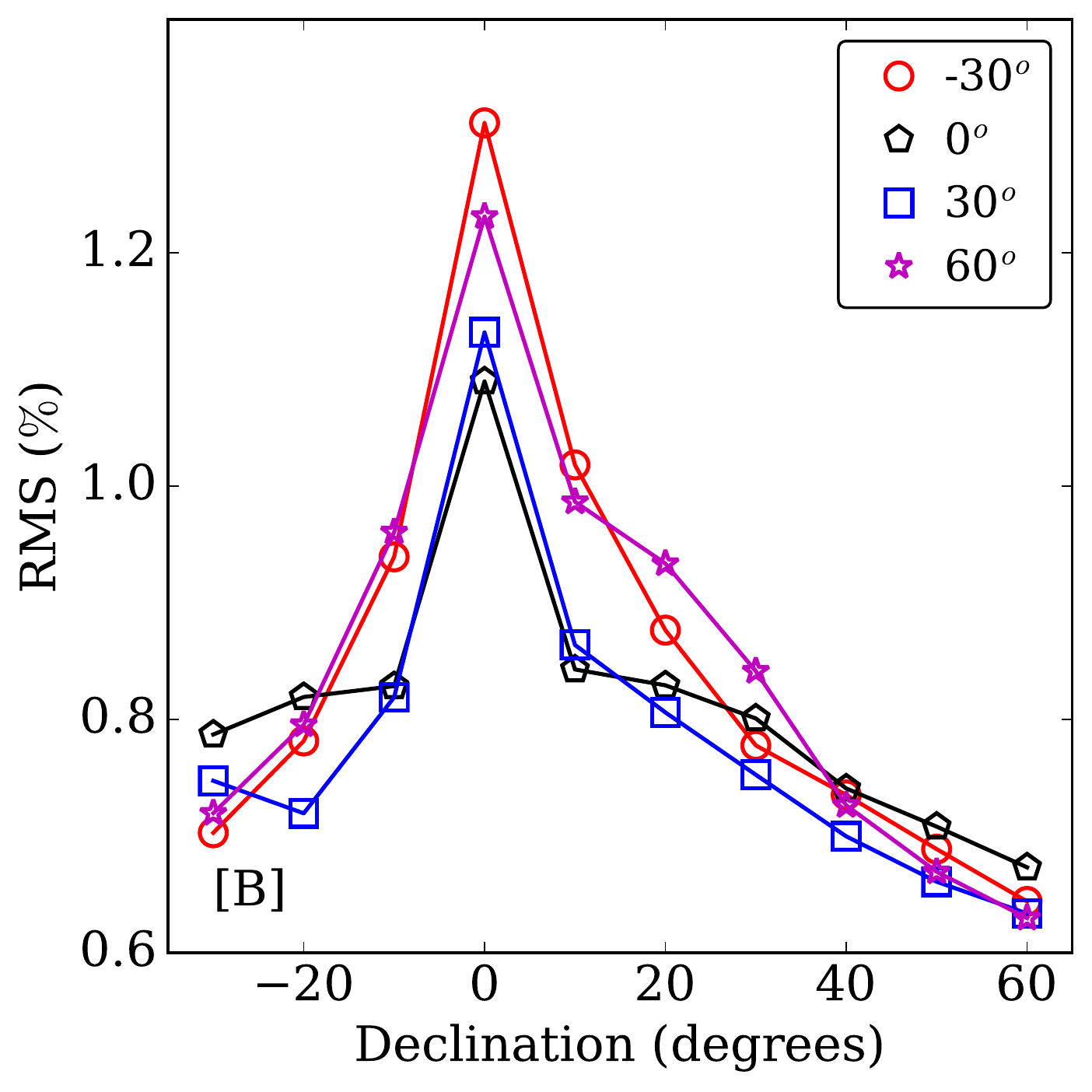} & 
\includegraphics[scale=0.375,trim={0.25cm 0.0cm 0.0cm 0.0cm},clip]{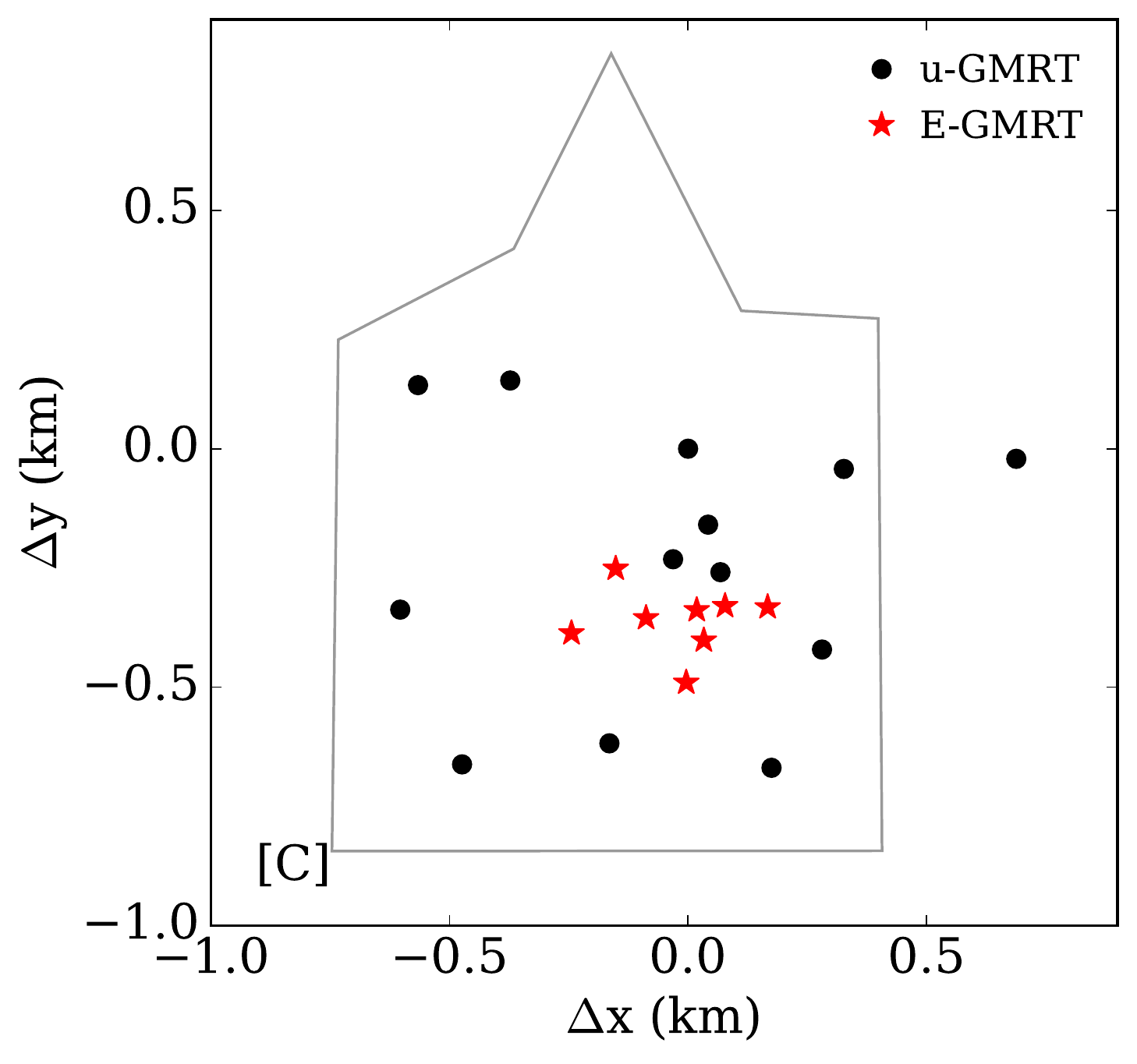} \\
\end{tabular}
\end{center}
\caption{Results for the optimization for FWHM~$=0.5$~km, i.e. $b_{\rm max}=1.0$~km.
	[A]~The Residual RMS plotted versus the number of new antennas, for $\delta = -30^\circ$, $0^\circ$, 
	$+30^\circ$, and $+60^\circ$; the dashed vertical line indicates 8 antennas, beyond which the 
	decline in Residual RMS with added antennas flattens out.
	[B]~The Residual RMS for the best configurations for the four declinations with 8 new antennas plotted
	against declination. The configuration with $\delta = +30^\circ$ yields the best overall performance.
	[C]~The locations of the new antennas, indicated by red stars, and of the existing GMRT antennas, indicated
	by solid black circles. See text for discussion.}
\label{fig:rms1}
\end{figure*}

\begin{figure*}
\begin{center}
\includegraphics[height=3.5in]{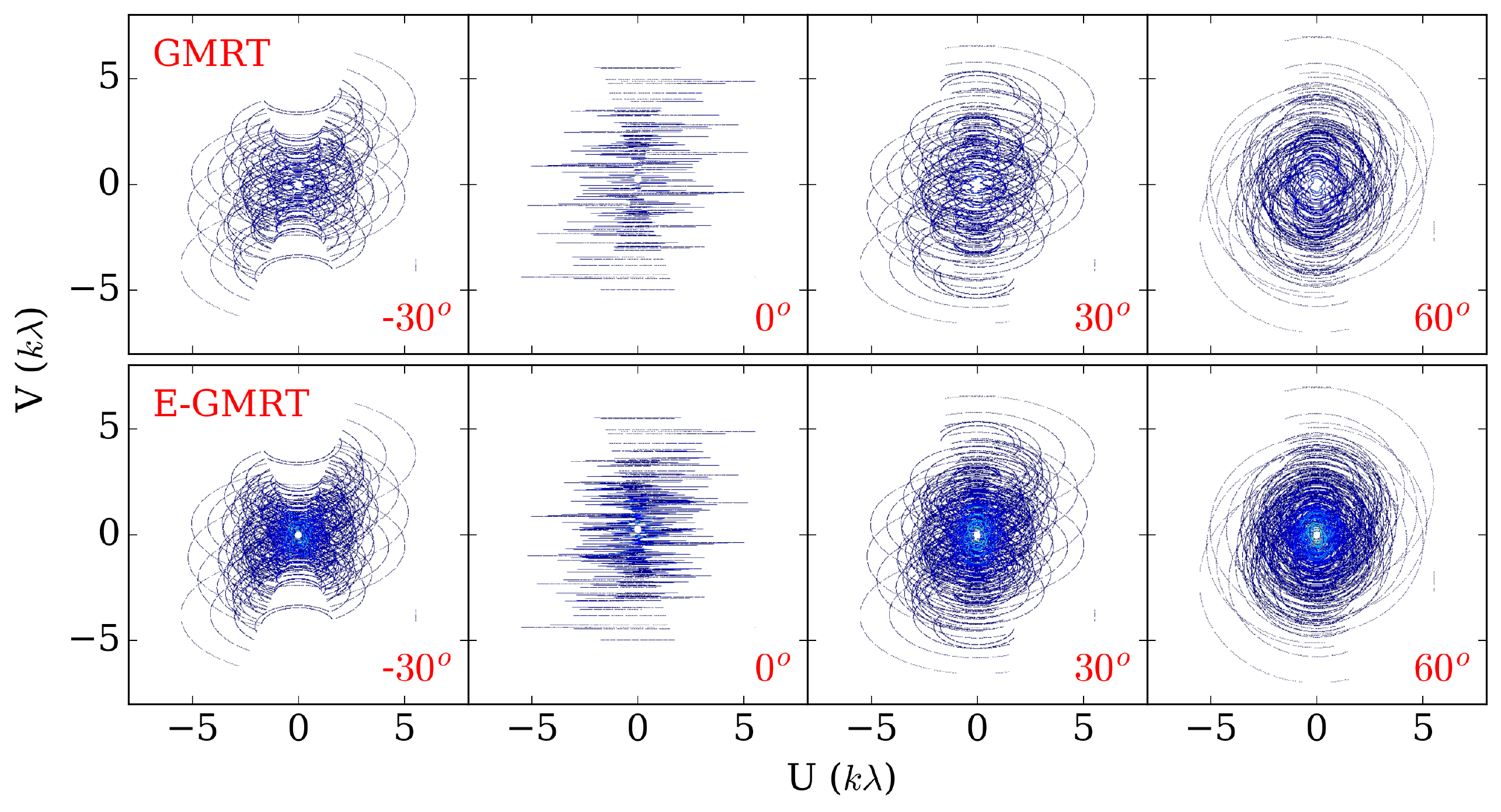}
\end{center}
\caption{A comparison between the U-V coverage of the uGMRT and the EGMRT, for baselines out to 
1~km (i.e. for the EGMRT optimization for FWHM~$=0.5$~km), for declinations~$\delta = -30^\circ$, 
$0^\circ$, $+30^\circ$, and $+60^\circ$, for a full-synthesis observing run at 1.2~GHz. The 
U-V coverage has been computed using a single channel.}
\label{fig:cov-full1}
\end{figure*}


Our optimization strategy was based on the fact that it is desirable for imaging of complex 
fields (e.g. the Galactic plane) to have a well-behaved synthesized beam over a range of 
angular resolutions, and especially at the shorter baselines which are sensitive to extended 
radio emission on a range of angular scales. We hence adopted an ``inside-out'' optimization 
strategy, first optimizing the array configuration for the shortest baselines (FWHM of the 
2-D Gaussian in the U-V plane of 0.5~km), then for short baselines (FWHM~$= 1.7$~km), then for 
intermediate baselines (FWHM~$= 5$~km), then for longer baselines (FWHM~$= 15$~km), and finally 
for the full array (FWHM~$=25$~km). In other words, instead of optimizing the full EGMRT configuration 
at once, we optimized the array configuration in steps, in which the optimization at each step 
is carried out for a given maximum baseline. At the next step, the antenna locations optimized 
in the previous step are kept fixed, and only the new added antenna locations are allowed to 
vary in the optimization. The resulting array would thus have a well-behaved synthesized beam 
over a range of angular resolutions, and not merely at the highest angular resolution.

We further note that the above FWHM's of the 2-D Gaussian distributions of the U-V coverage were not set to 
be equal to the longest baseline of the array whose configuration was being optimized. This was done 
because there are, of course, baselines beyond the FWHM that contribute to the 2-D Gaussian U-V distribution. 
We hence allowed longest baselines of $b_{\rm max} \approx 1$~km for FWHM~$= 0.5$~km, of $\approx 5$~km for 
FWHM~$\approx 1.7$~km, of $\approx 15$~km for FWHM~$\approx 5$~km, of $\approx 25$~km for 
FWHM~$\approx 15$~km, of $\approx 50$~km for FWHM $\approx 25$~km. 

As mentioned above, the optimizations out to $b_{\rm max} = 5.0$~km (i.e. FWHM~$\approx 1.7$~km) were carried out 
using the random sampling approach, as the total number of possible antenna locations is relatively small. 
Since the GMRT antennas have a diameter of 45~m, we divided the possible antenna locations (including any 
land constraints) into cells of size $60$~m $\times 60$~m (smaller cells would have meant large shadowing 
of antennas by each other; note that our optimization approach does not include a penalty for shadowing). 
For a region of size $\approx 1.7$~km~$\times 1.7$~km, this meant $\approx 800$ possible locations for 
the new antennas, after including the land constraints. This could be handled in reasonable computing time 
via random sampling, estimating the Residual RMS for $10^4$ random antenna configurations, and then choosing 
the configuration with the minimum Residual RMS.  We hence chose to use random sampling for maximum baselines 
out to $b_{\rm max} = 5$~km. We also verified that very similar antenna configurations were obtained using 
{\sc iAntConfig} and random-sampling for the most compact configuration, with FWHM~$\approx 0.5$~km. For 
$b_{\rm max} \gg 5$~km, it was clear that random sampling would not 
provide sufficient coverage of the possible antenna configurations in reasonable computing time. For the 
optimization for $b_{\rm max} \geq 15$~km (i.e. FWHM~$\gtrsim 5$~km), we hence used the tomographic projection 
approach \citep{devilliers07}.

\begin{figure*}
\begin{center}
	\begin{tabular}{ccc}
	\includegraphics[scale=0.375,trim={0.25cm 0.0cm 0.0cm 0.0cm},clip]{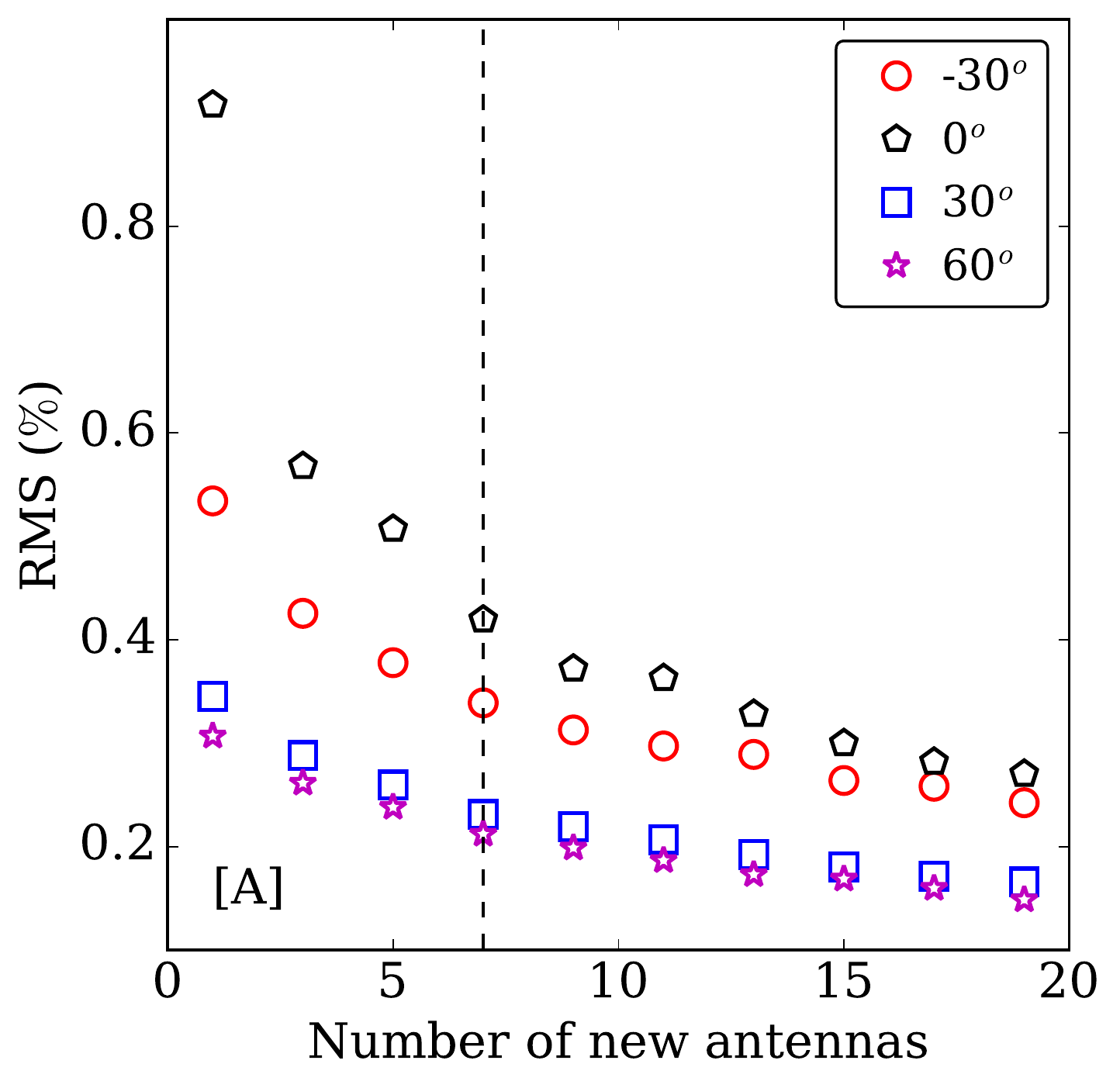} & 
	\includegraphics[scale=0.375,trim={0.25cm 0.0cm 0.0cm 0.0cm},clip]{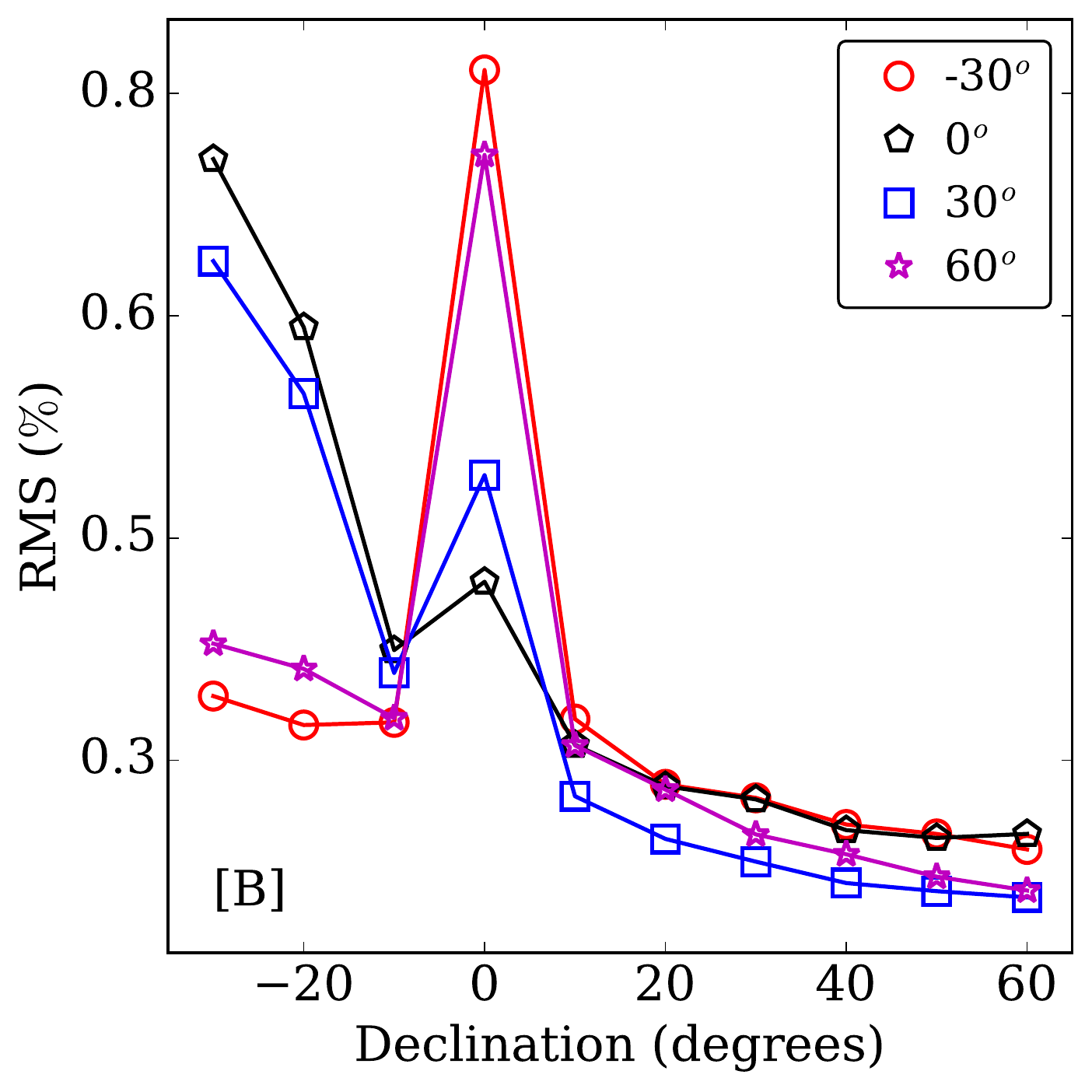} & 
	\includegraphics[scale=0.375,trim={0.25cm 0.0cm 0.0cm 0.0cm},clip]{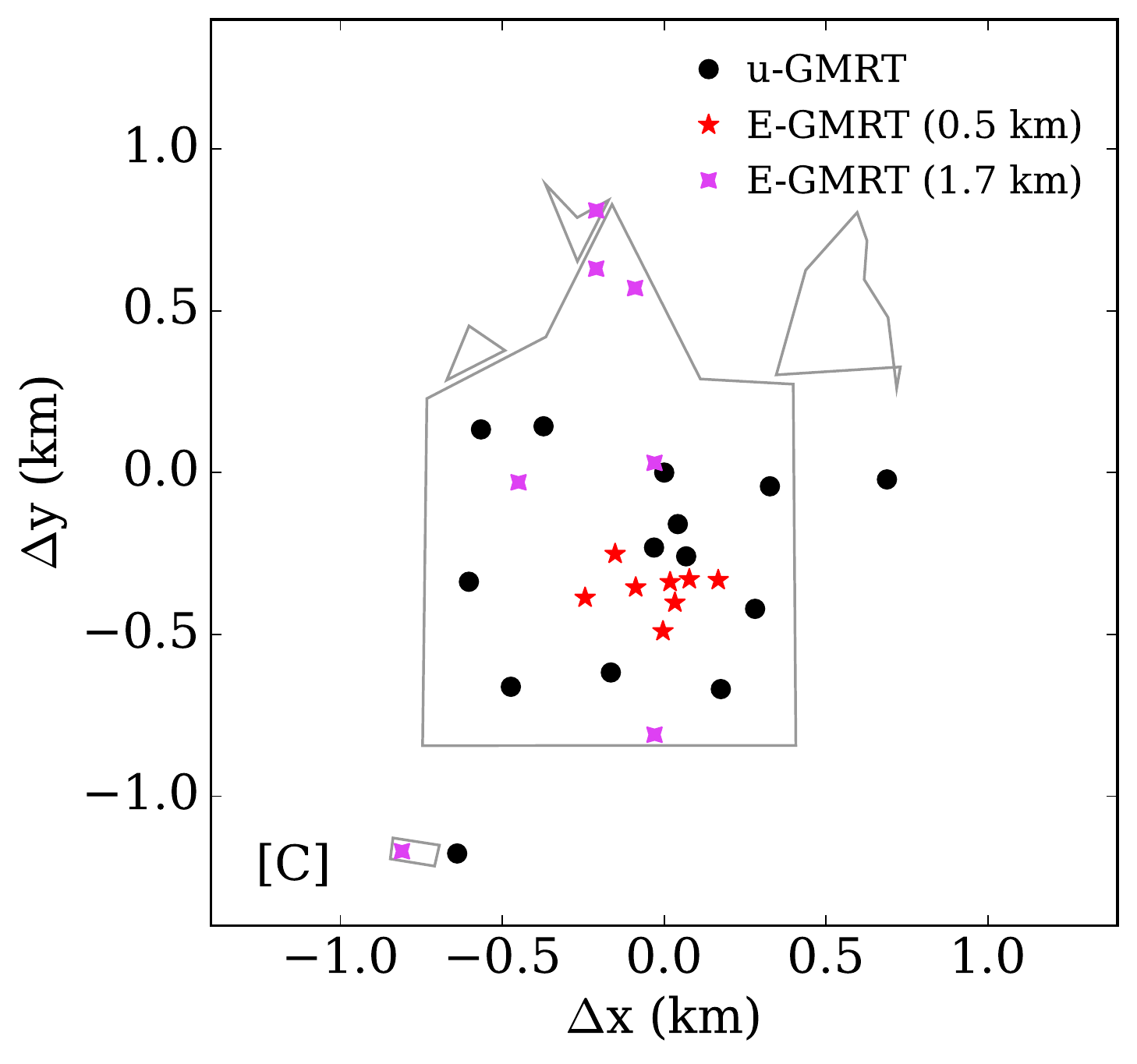} \\
\end{tabular}
\end{center}
\caption{Results for the optimization for FWHM~$=1.7$~km, i.e. $b_{\rm max}=5$~km.
	[A]~The Residual RMS plotted versus the number of new antennas, for $\delta = -30^\circ$, $0^\circ$, 
	$+30^\circ$, and $+60^\circ$; the dashed vertical line indicates 7 antennas, beyond which the 
	decline in Residual RMS with added antennas flattens out.
	[B]~The Residual RMS for the best configurations for the four declinations with 7 new antennas plotted
	against declination. The configuration with $\delta = +30^\circ$ yields the best overall performance.
	[C]~The locations of the GMRT antennas are indicated by solid black circles, of the 8 new antennas 
	obtained in the FWHM~$=0.5$~km optimization by red stars, and of the 7 new antennas obtained here by 
	magenta stars. See text for discussion.}
\label{fig:rms2}
\end{figure*}

\begin{figure*}
\begin{center}
\includegraphics[height=3.5in]{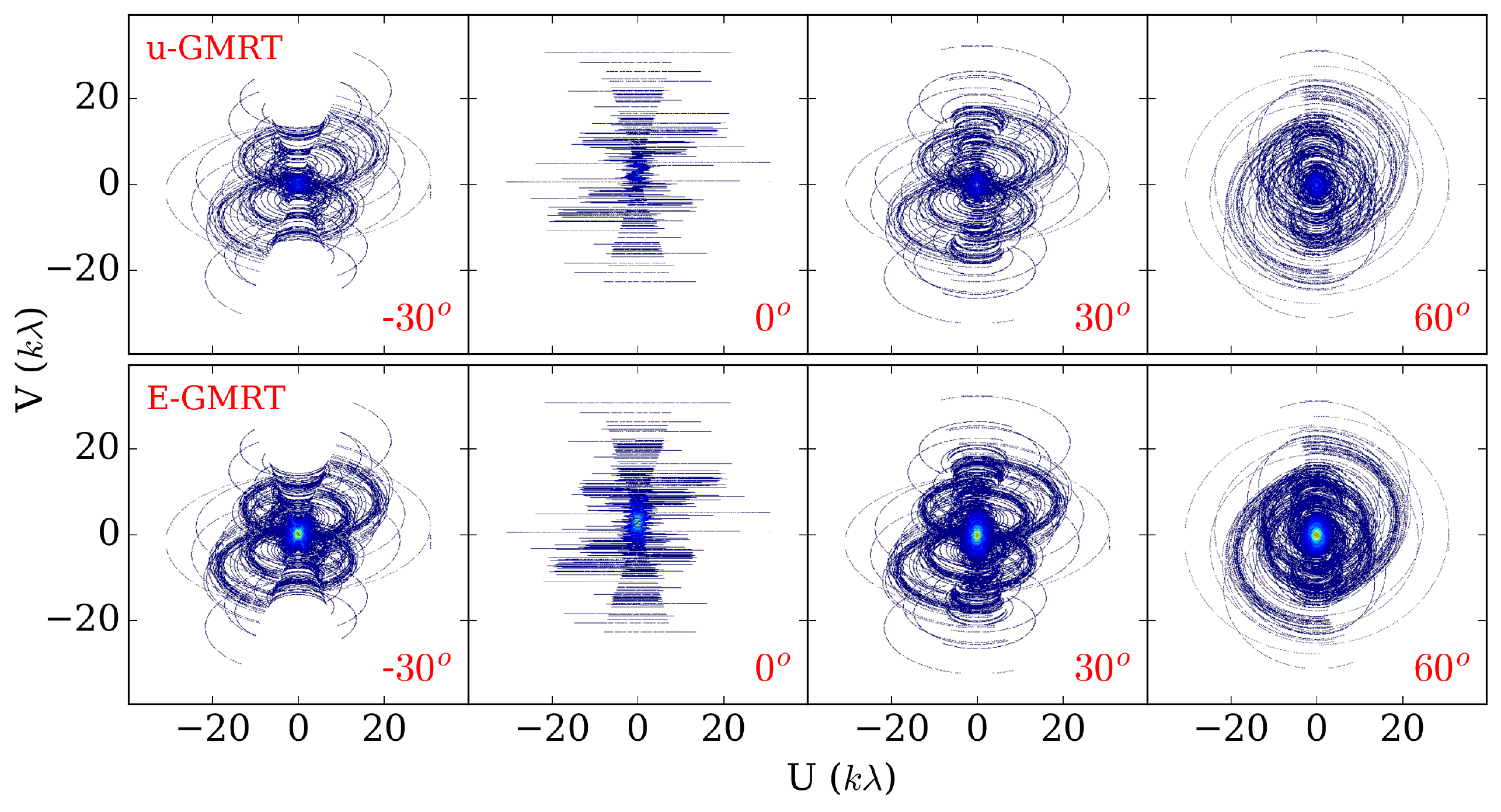}
\end{center}
\caption{A comparison between the single-channel U-V coverage of the GMRT and the EGMRT, for baselines out to 
5~km (i.e. for the EGMRT optimization for FWHM~$= 1.7$~km), for the four declinations, for a full 
synthesis observing run at 1.2~GHz.}
\label{fig:cov-full2}
\end{figure*}


\begin{figure*}
\begin{center}
	\begin{tabular}{ccc}
	\includegraphics[scale=0.375,trim={0.25cm 0.0cm 0.0cm 0.0cm},clip]{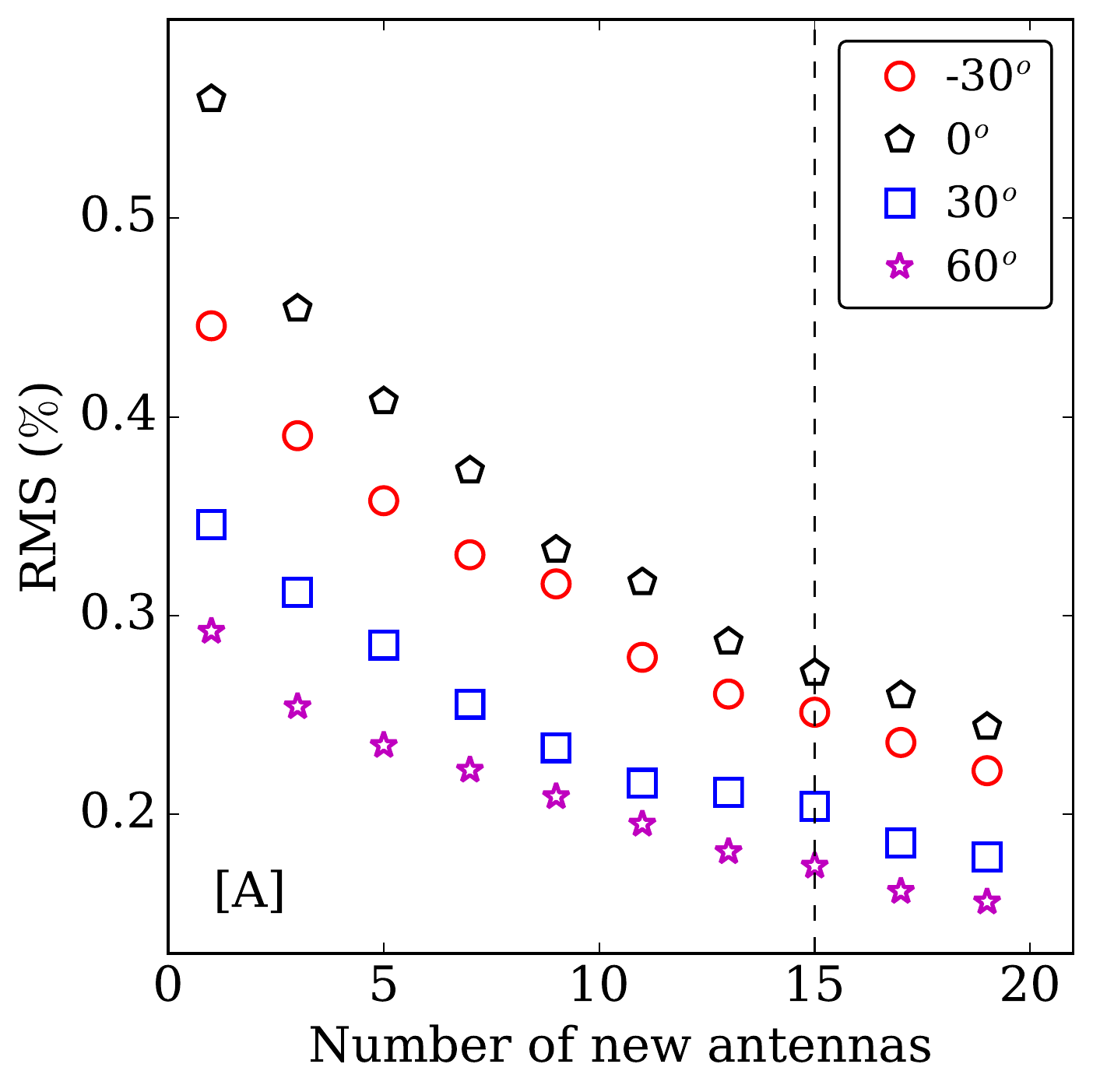} & 
	\includegraphics[scale=0.375,trim={0.25cm 0.0cm 0.0cm 0.0cm},clip]{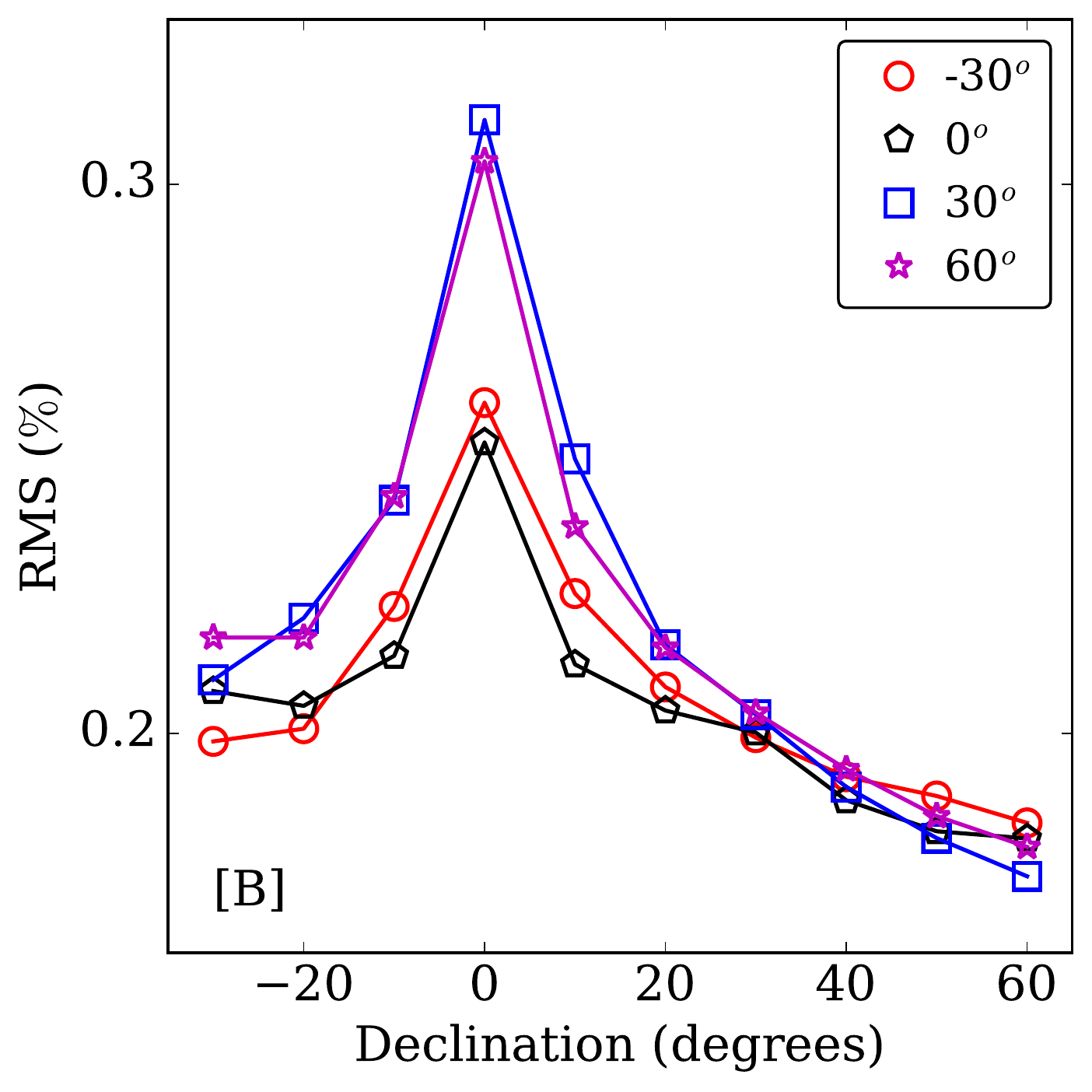} & 
        \includegraphics[scale=0.375,trim={0.25cm 0.0cm 0.0cm 0.0cm},clip]{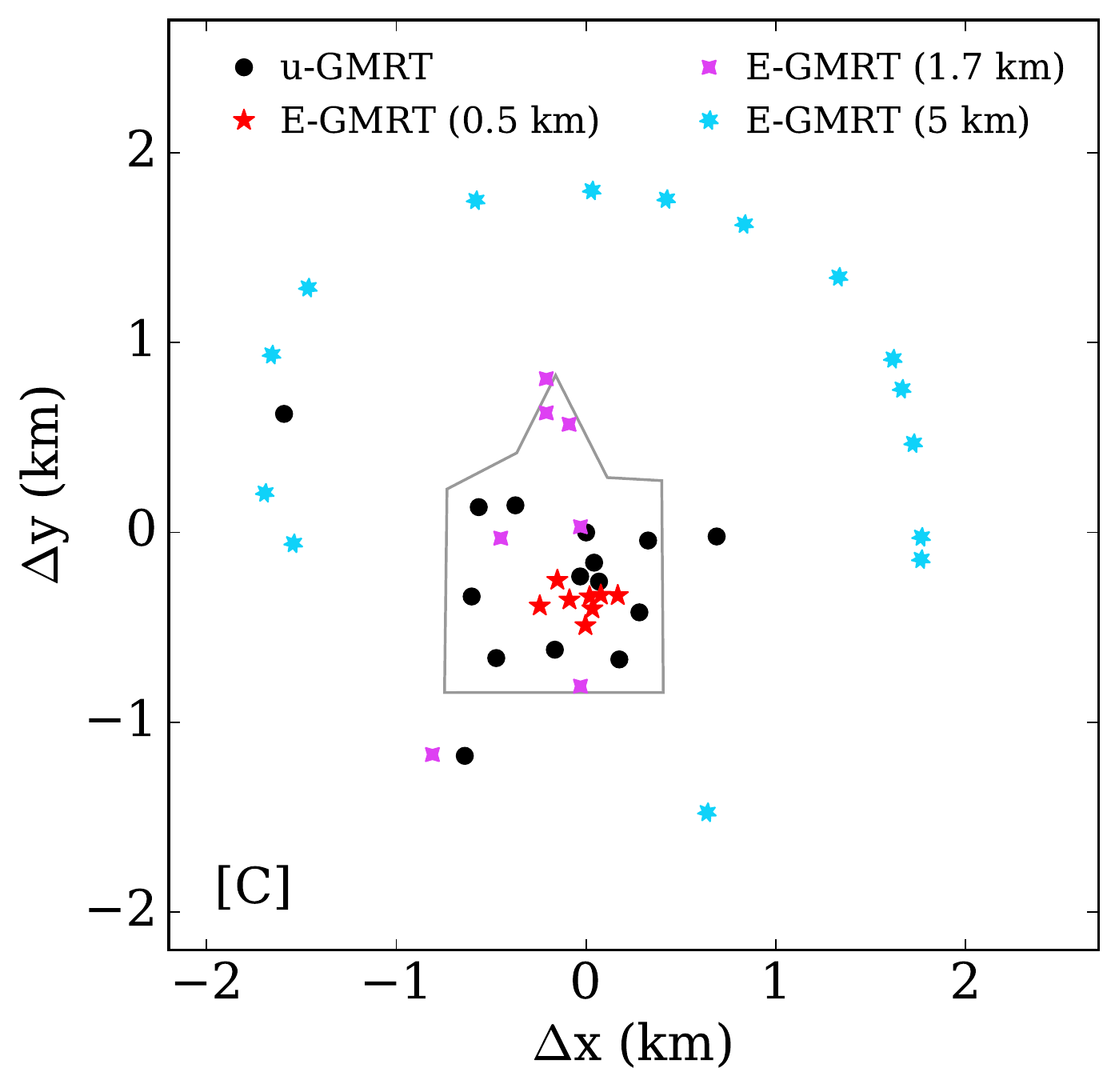} \\
\end{tabular}
\end{center}
\caption{Results for the optimization for FWHM~$=5$~km, i.e. $b_{\rm max}=15$~km.
	[A]~The Residual RMS plotted versus the number of new antennas, for $\delta = -30^\circ$, $0^\circ$, 
	$+30^\circ$, and $+60^\circ$; the dashed vertical line indicates 15 antennas, beyond which the 
	decline in Residual RMS with added antennas slows down. 
	[B]~The Residual RMS for the best configurations for the four declinations with 15 new antennas plotted
	against declination. The configuration with $\delta = 0^\circ$ yields the best overall performance.
	[C]~The locations of the GMRT antennas are indicated by solid black circles, of the 8 new antennas 
	obtained in the FWHM~$=0.5$~km optimization by red stars, of the 7 new antennas obtained in the FWHM~$=1.7$~km 
	optimization by magenta stars, and of the 15 new antennas obtained here by blue stars. See text for discussion.}
\label{fig:rms3}
\end{figure*}

\begin{figure*}
\begin{center}
\includegraphics[height=3.5in]{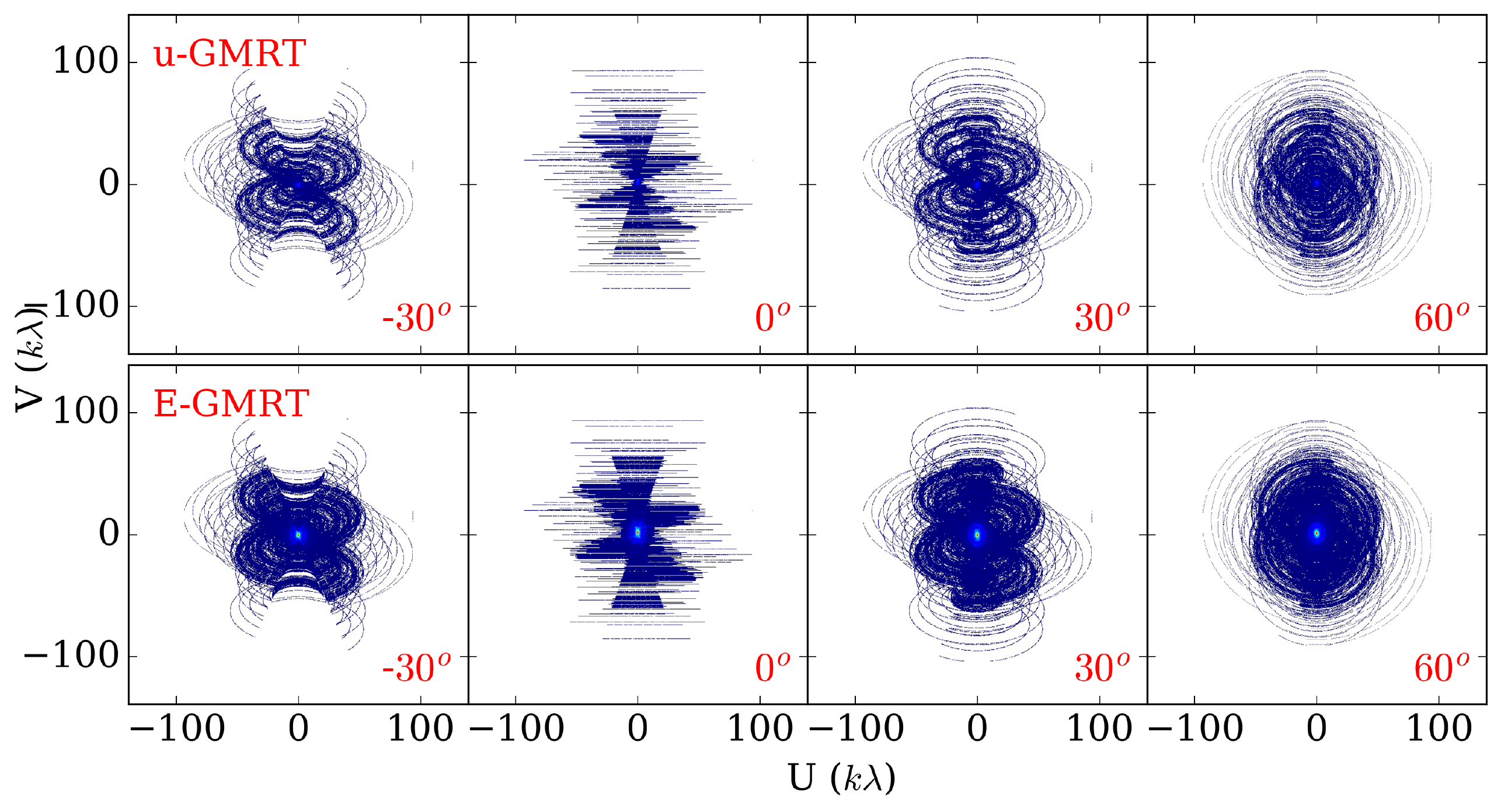}
\caption{A comparison between the single-channel U-V coverage of the GMRT and the EGMRT, for a maximum baseline 
of 15~km (i.e. for the EGMRT optimization for FWHM~$=5.0$~km), for the four declinations, for a full 
synthesis observing run at 1.2~GHz.
	\label{fig:cov-full3}}
\end{center}
\end{figure*}


\subsubsection{Optimization for FWHM~$= 0.5$~km, i.e. $b_{\rm max} = 1.0$~km}
\label{sec:bmax0.5km}

The first step of our optimization is for FWHM~$= 0.5$~km. Here, as noted above, we used both the 
random sampling and tomographic projection methods, and obtained very similar results from the two approaches.
Both approaches are described in detail below. We finally used the results from the tomographic projection method. 

The spatial density of the existing GMRT antennas is maximum in the central square, with baselines of 
$\lesssim 1$~km. Hence, the optimization for FWHM~$=0.5$~km, following both approaches, was carried out 
by restricting the EGMRT antenna locations to lie within the central square. Further, we included only 
the 14 GMRT central square antennas as ``fixed'' antennas in the optimization.  

In the tomographic projection method, beginning with the above fixed GMRT antennas, we add N new antennas 
to the array, starting with ${\rm N}=1$, and increasing the number of new antennas by one at each 
iteration. For each value of N, we carried out the tomographic projection optimization for 100 random initial 
conditions for the new antenna locations, and evaluated the final Residual RMS for each of the 100 realizations. 
We then chose the realization with the lowest Residual RMS as the optimal configuration for the N new antennas. 
This minimum Residual RMS was saved, along with the locations of the new antennas, and the process repeated for 
(N+1) new antennas. The Residual RMS initially declines steeply with each new added antenna, as the U-V coverage 
approaches a 2-D Gaussian U-V distribution. However, beyond some number of new antennas, there is no significant 
decrease in the Residual RMS, i.e. improvement in the U-V coverage, on adding more antennas. We fixed the 
number of new antennas to the number above which the Residual RMS does not change by more than $\approx 10$\% 
with the addition of further new antennas.

In the random-sampling method, we again add N new antennas to the array, starting with ${\rm N}=1$, 
and increasing the number of new antennas by one at each iteration. However, here we randomly assign the new antennas 
to available cells in the GMRT central square, keeping the locations of the existing GMRT antennas fixed. We then 
evaluate the Residual RMS for the configuration, comparing the U-V coverage of the configuration with the ``ideal'' 
2-D Gaussian of FWHM~$=0.5$~km. We carry out this random assignment process $10^4$ times for each value of N, and 
determine the minimum Residual RMS for the $10^4$ evaluated configurations. This minimum Residual RMS is then stored, 
again along with the locations of the new antennas, and the process is repeated, by adding (N+1) antennas. We find a 
pattern similar to that seen in the tomographic projection approach, i.e. that the Residual RMS initially declines steeply
with each added antenna, but, beyond some number of new antennas, does not decrease significantly on adding more 
antennas. Again, the number of new antennas is then fixed to the number above which the Residual RMS does not 
change by more than $\approx 10$\% with the addition of further new antennas.

As noted earlier, the above process was carried out independently for four declinations, $\delta = -30^\circ$, $+30^\circ$ 
and $+60^\circ$. Fig.~\ref{fig:rms1}[A] shows the Residual RMS plotted against the number of antennas for the four 
different declinations, with the results obtained via the tomographic projection method. The dashed vertical line 
indicates 8 added antennas, beyond which the Residual RMS improves only slowly with additional antennas (for 
all four declinations).

Following this, the Residual RMS obtained from the four ``best'' configurations with 8 EGMRT antennas (from each of the 
four declinations) was then evaluated as a function of declination, over the declination range $-30^\circ$ to $+60^\circ$, 
to identify the best configuration for all declinations. Fig.~\ref{fig:rms1}[B] shows the Residual RMS plotted 
versus declination for the four different configurations, labelled by the declination at which the configuration was 
optimized. For the case of FWHM~$=0.5$~km, we find that the best results over the full declination range were 
obtained from the configuration optimized for $\delta = +30^\circ$. This array configuration is shown in Fig.~\ref{fig:rms1}[C],
with the 14 GMRT central square antennas shown as solid black circles and the 8 new EGMRT antennas shown as 
red stars.

Finally, Fig.~\ref{fig:cov-full1} compares the U-V coverage obtained in a full-synthesis observing run with 
the 14 GMRT central square antennas with that obtained with the EGMRT array with 8 new antennas, for 
$\delta = -30^\circ$, $0^\circ$, $+30^\circ$, and $+60^\circ$. It is clear that the EGMRT array would 
yield significantly better U-V coverage than that of the present GMRT for all declinations; similar 
results are obtained for snapshot observations. 


\begin{figure*}
\begin{center}
	\begin{tabular}{ccc}
	\includegraphics[scale=0.375,trim={0.25cm 0.0cm 0.0cm 0.0cm},clip]{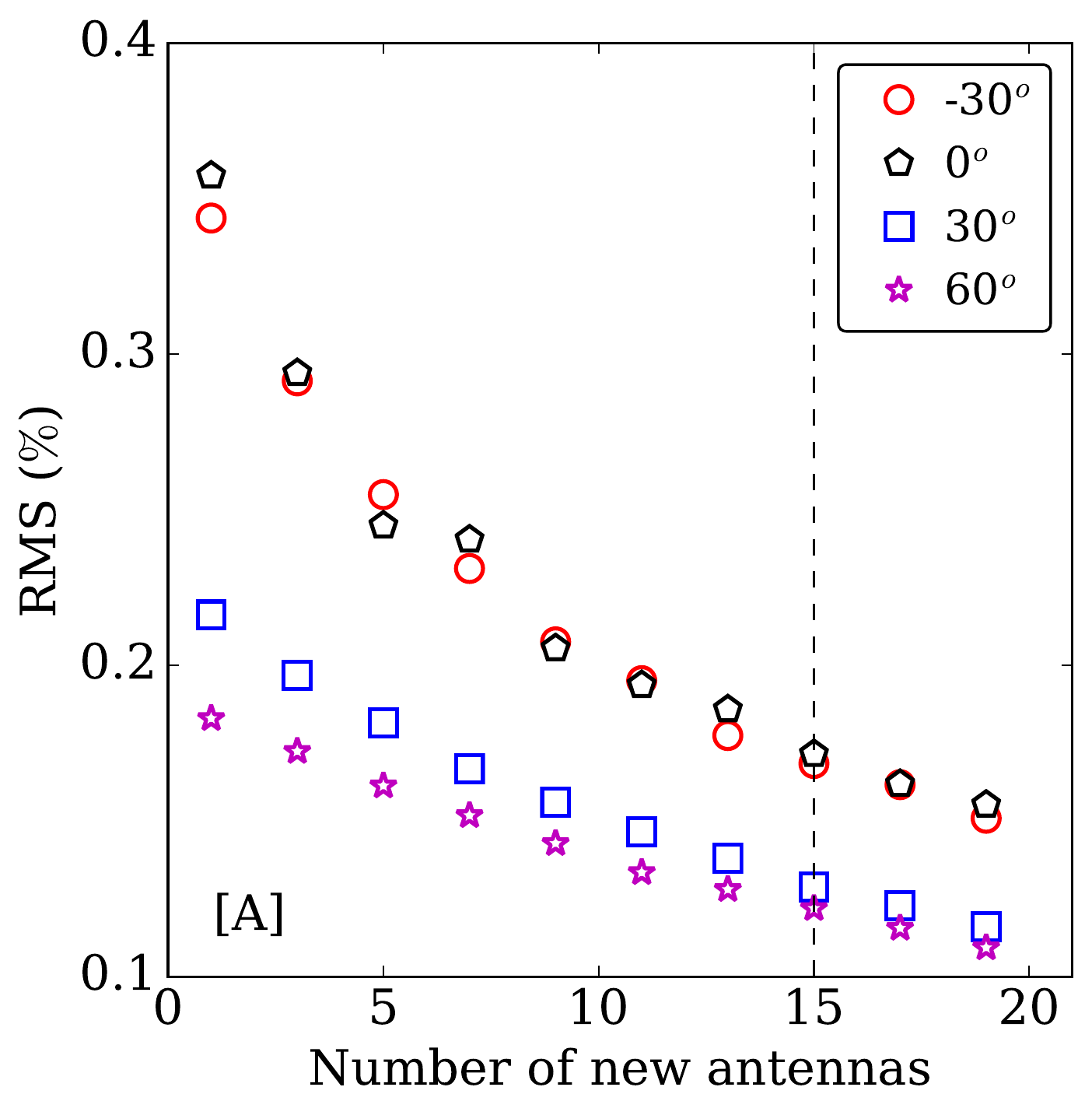}  &
	\includegraphics[scale=0.375,trim={0.25cm 0.0cm 0.0cm 0.0cm},clip]{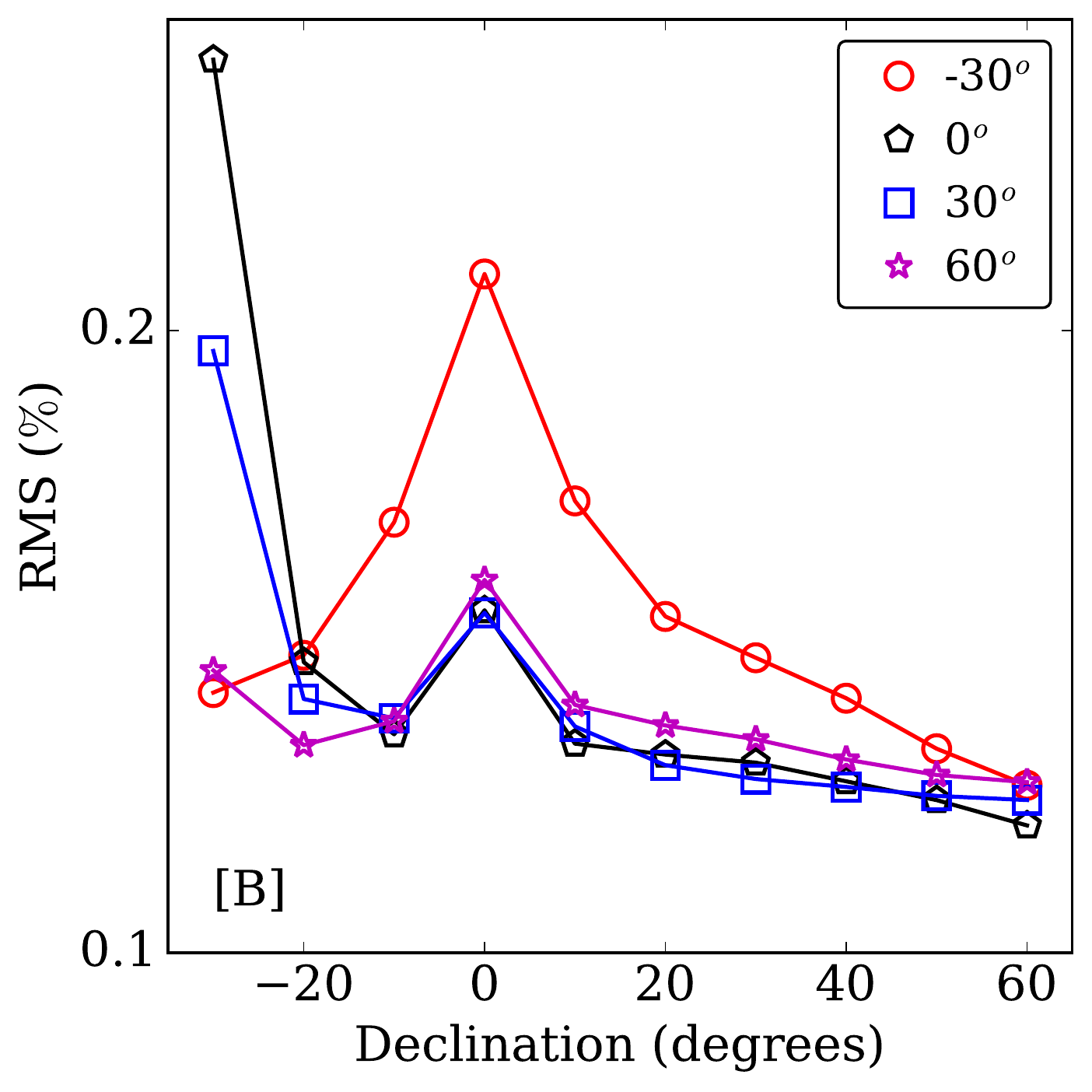} &
	\includegraphics[scale=0.375,trim={0.25cm 0.0cm 0.0cm 0.0cm},clip]{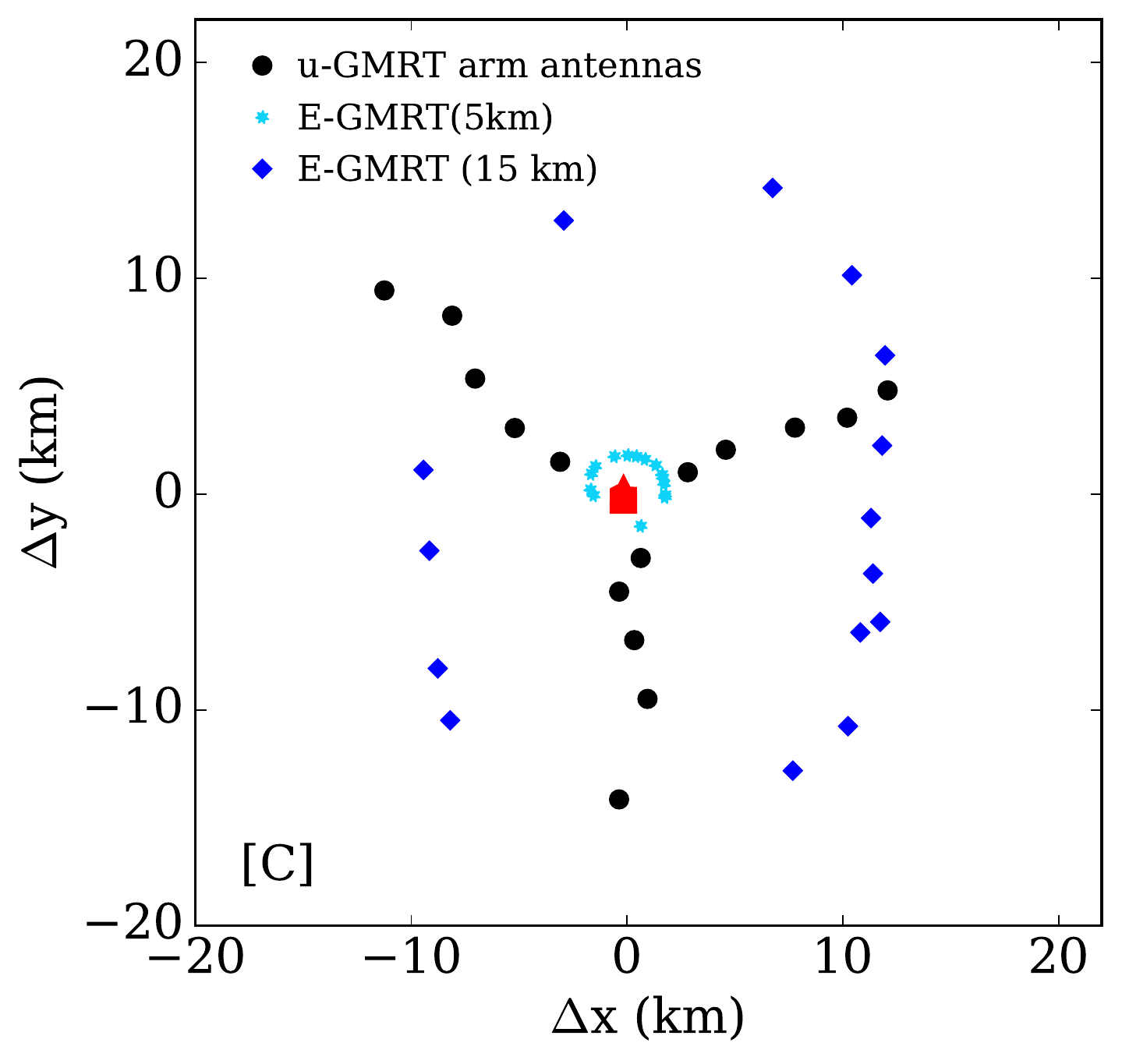} \\
\end{tabular}
\caption{Results for the optimization for FWHM~$=15$~km, i.e. $b_{\rm max}=25$~km.
[A]~The Residual RMS plotted versus the number of new antennas, for $\delta = -30^\circ$, $0^\circ$, 
$+30^\circ$, and $+60^\circ$; the dashed vertical line indicates 15 antennas, beyond which the 
decline in Residual RMS with added antennas slows down. 
[B]~The Residual RMS for the best configurations for the four declinations with 15 new antennas plotted
against declination. The configuration with $\delta = +60^\circ$ yields the best overall performance.
[C]~The locations of the GMRT antennas are indicated by solid black circles, of the 15 new antennas obtained 
in the FWHM~$=5.0$~km optimization by cyan stars, and of the 15 new antennas obtained here by 
blue diamonds. See text for discussion.
\label{fig:rms4}}
\end{center}
\end{figure*}

\begin{figure*}
\begin{center}
\includegraphics[height=3.5in]{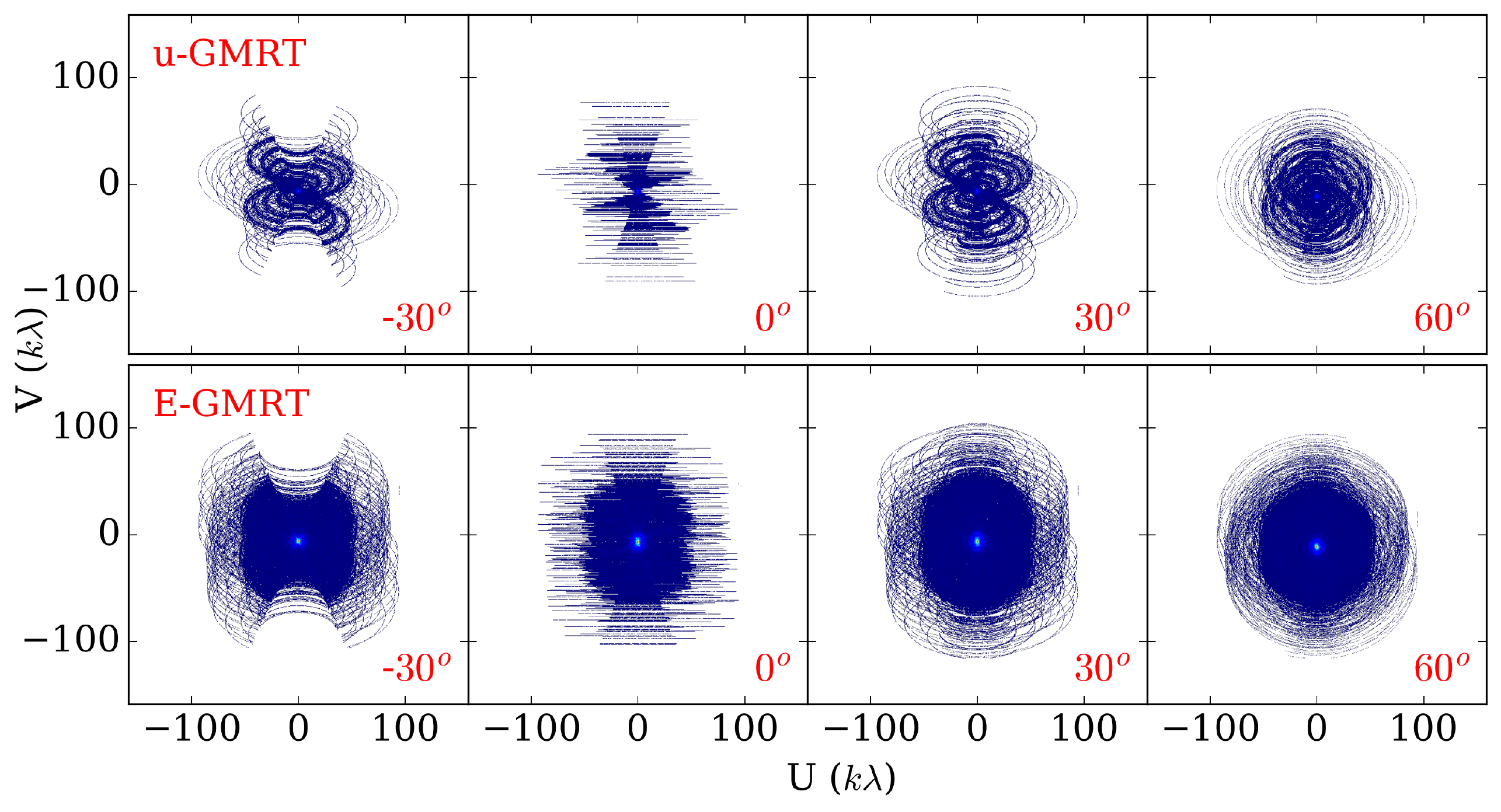}
\caption{A comparison between the single-channel U-V coverage of the GMRT and the EGMRT, for baselines out to 
25~km (i.e. for the EGMRT optimization for FWHM~$=15$~km), for the four declinations, for a full 
synthesis observing run at 1.2~GHz.
	\label{fig:cov-full4}}
\end{center}
\end{figure*}


\subsubsection{Optimization for FWHM~$= 1.7$~km, i.e. $b_{\rm max} = 5.0$~km}
\label{sec:bmax5km}

For the next step, the optimization for FWHM~$=1.7$~km, we included all existing GMRT antennas with baselines out 
to 5~km (twenty antennas in all), as well as the 8 new EGMRT antenna locations obtained above, as fixed antennas. The 
new antenna locations were constrained to be on land either currently owned by the GMRT or on land that might be acquired 
for the expansion.  We note that this implied significant constraints on the allowed antenna locations, and hence, that 
better results (i.e. a lower Residual RMS) were obtained with the random-sampling method than with the tomographic 
projection method (which applies land constraints at the end of the optimization in an {\it ad~hoc} manner). We hence used 
the random-sampling approach to identify the new antenna locations for 
$b_{\rm max} = 5.0$~km, following the procedure described in Section~\ref{sec:bmax0.5km}, except that new antennas were 
added in steps of two, rather than one, to reduce the computational requirements.

Fig.~\ref{fig:rms2}[A] shows the Residual RMS plotted against the number of antennas for the four different declinations; 
the dashed vertical line is for 7 new antennas, above which we do not find a significant improvement in the Residual 
RMS on adding further antennas. Fig.~\ref{fig:rms2}[B] shows the Residual RMS plotted against declination for the 
best configurations obtained with 7 new antennas for the four different declinations; we find that the best overall 
performance is obtained for the configuration for $\delta = -30^\circ$. This is shown in Fig.~\ref{fig:rms2}[C], 
with GMRT antennas shown as solid black circles, EGMRT antennas obtained from the FWHM~$=0.5$~km optimization 
as red stars, and EGMRT antennas obtained from the FWHM~$=1.7$~km optimization as magenta stars. Fig.~\ref{fig:cov-full2} 
shows a comparison between the single-channel U-V coverage of the GMRT array with baselines out to $b_{\rm max} = 5.0$~km 
and the EGMRT array (with 20 GMRT antennas and 15 new antennas), for a full-synthesis observing run and the different 
declinations. Again, it is clear that significantly better U-V coverage is obtained with the EGMRT array.

\begin{figure*}
\begin{center}
	\begin{tabular}{ccc}
	\includegraphics[scale=0.375,trim={0.25cm 0.0cm 0.0cm 0.0cm},clip]{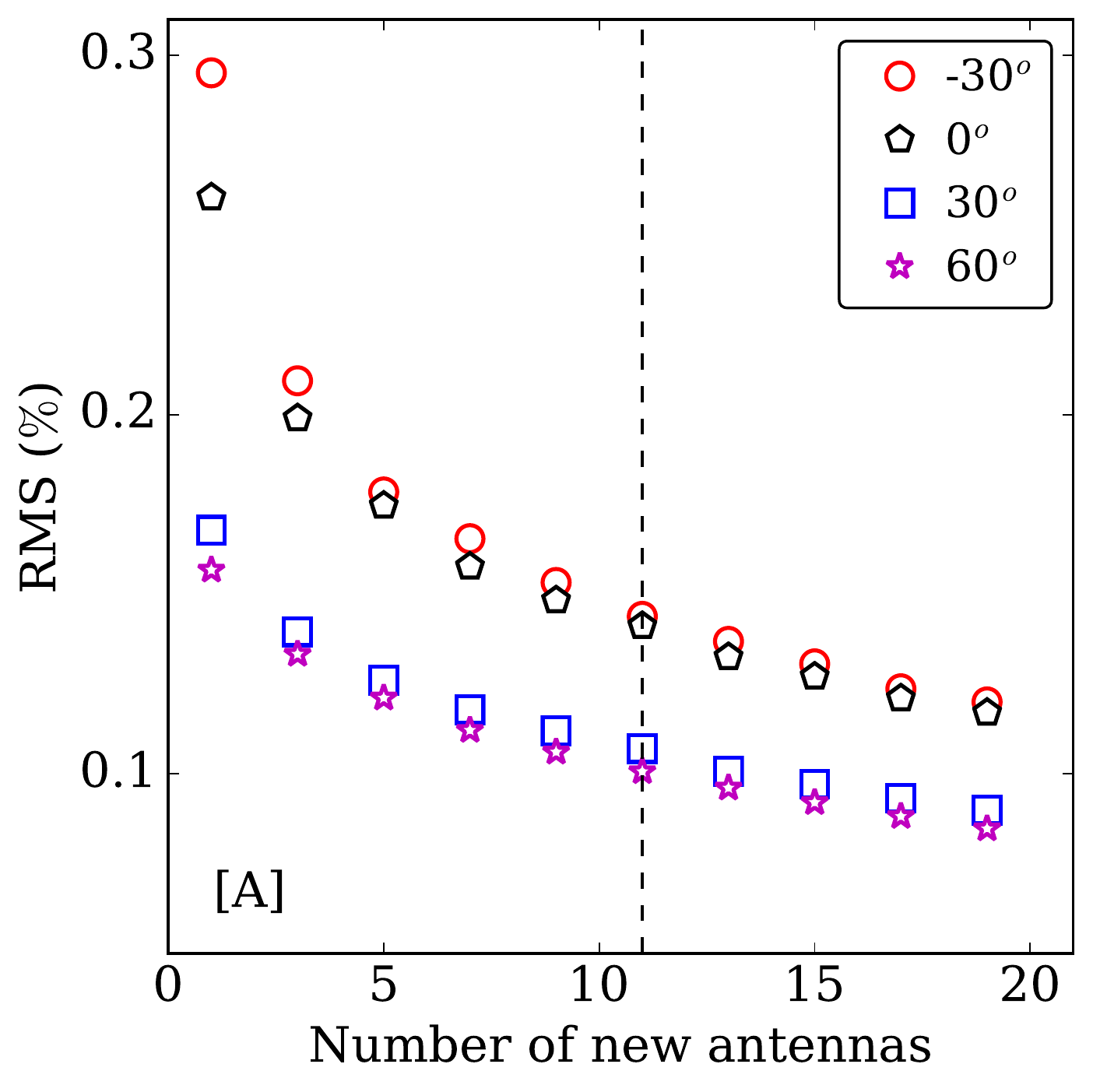}  & 
	\includegraphics[scale=0.375,trim={0.25cm 0.0cm 0.0cm 0.0cm},clip]{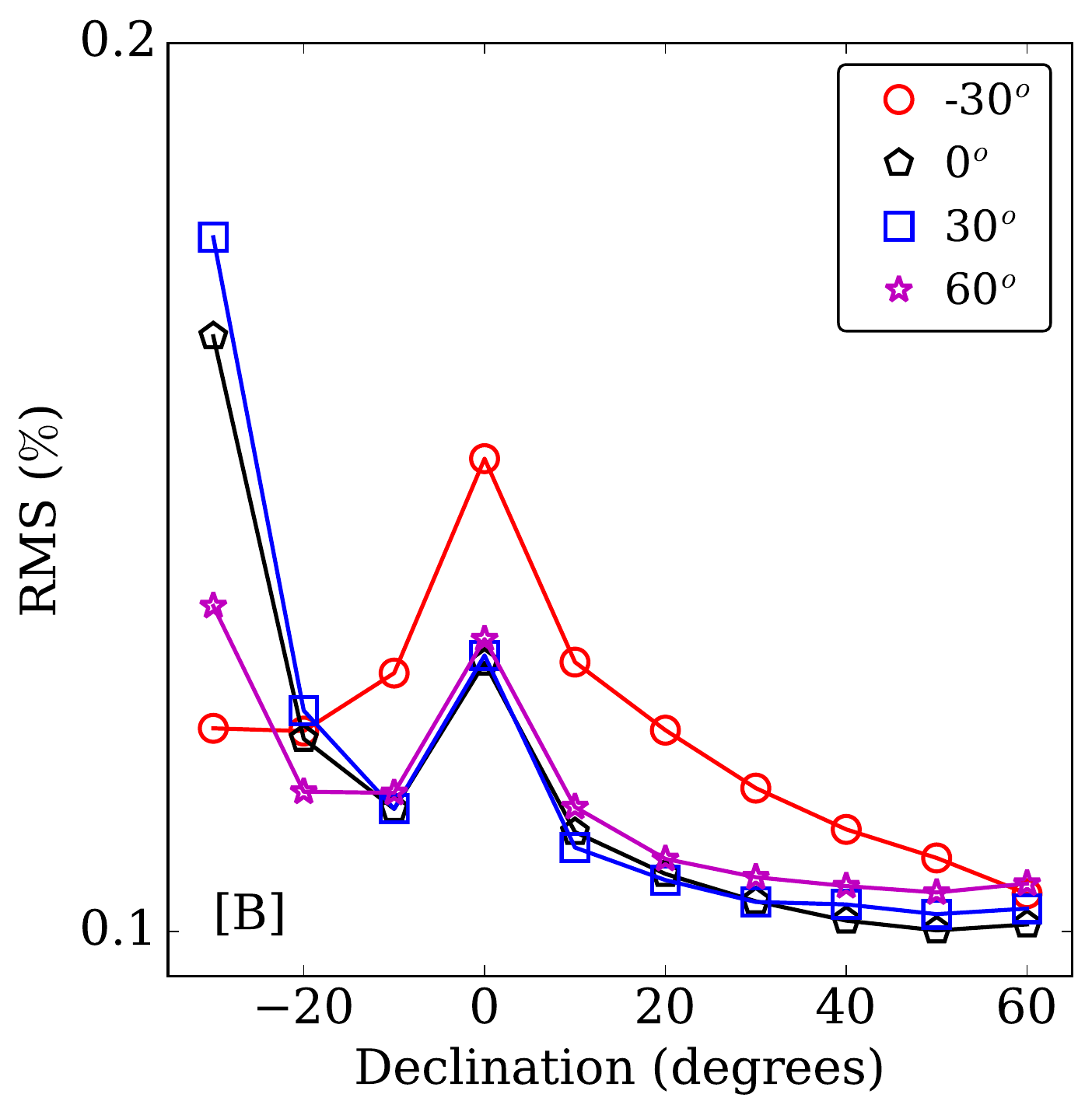} & 
	\includegraphics[scale=0.375,trim={0.25cm 0.0cm 0.0cm 0.0cm},clip]{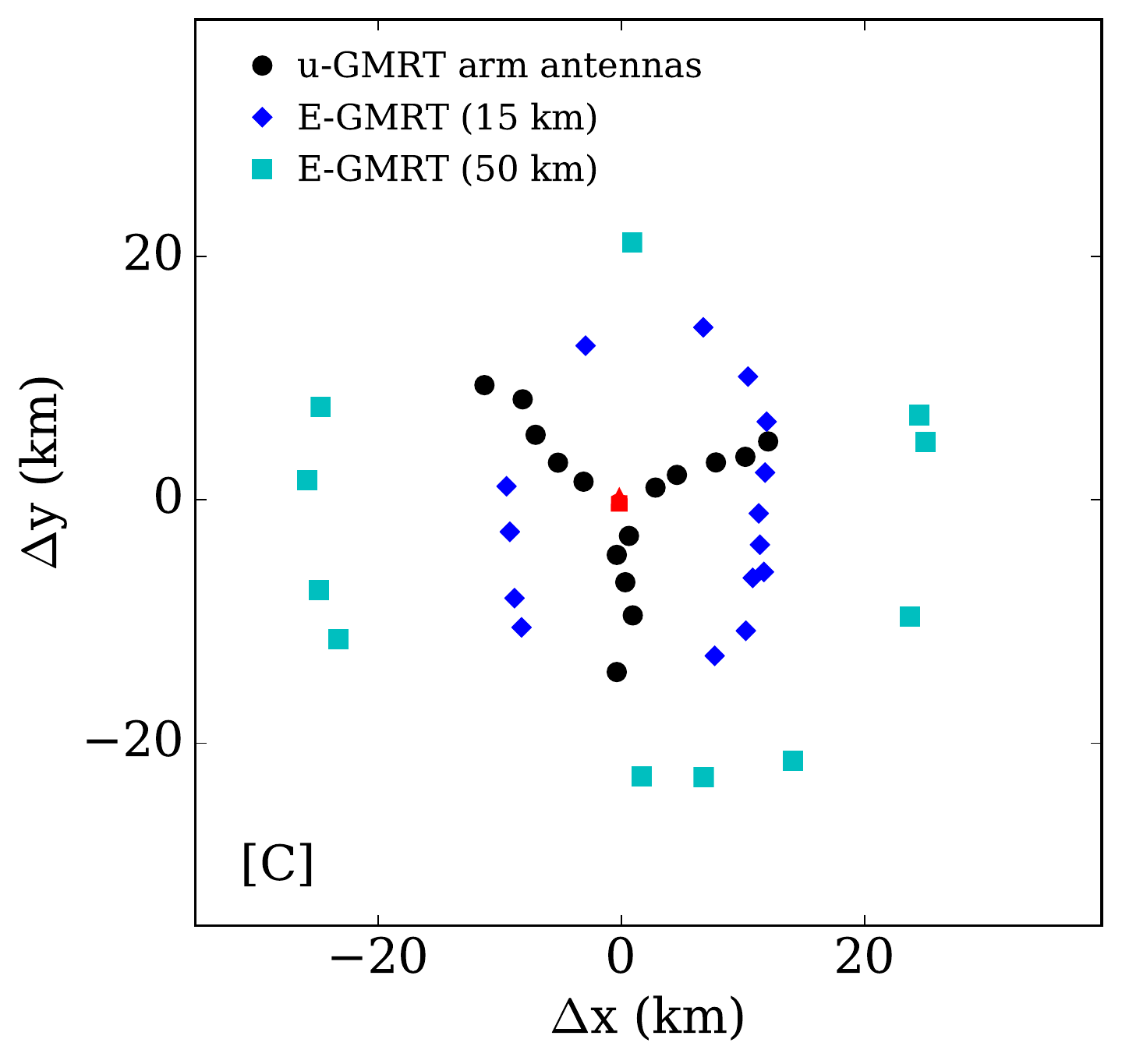} \\
\end{tabular}
\end{center}
	\caption{Results for the optimization for FWHM~$=25$~km, i.e. $b_{\rm max}=50$~km.
	[A]~The Residual RMS plotted versus the number of new antennas, for $\delta = -30^\circ$, $0^\circ$, 
	$+30^\circ$, and $+60^\circ$; the dashed vertical line indicates 11 antennas, beyond which the 
	decline in Residual RMS with added antennas slows down. 
	[B]~The Residual RMS for the best configurations for the four declinations with 11 new antennas plotted
	against declination. The configuration with $\delta = +60^\circ$ yields the best overall performance.
	[C]~The locations of the GMRT antennas are indicated by solid black circles, of the 15 new antennas obtained 
	in the FWHM~$=15.0$~km optimization by blue diamonds, and of the 11 new antennas obtained here by cyan
	squares. See text for discussion.}
\label{fig:rms5}
\end{figure*}

\begin{figure*}
\begin{center}
\includegraphics[height=3.5in]{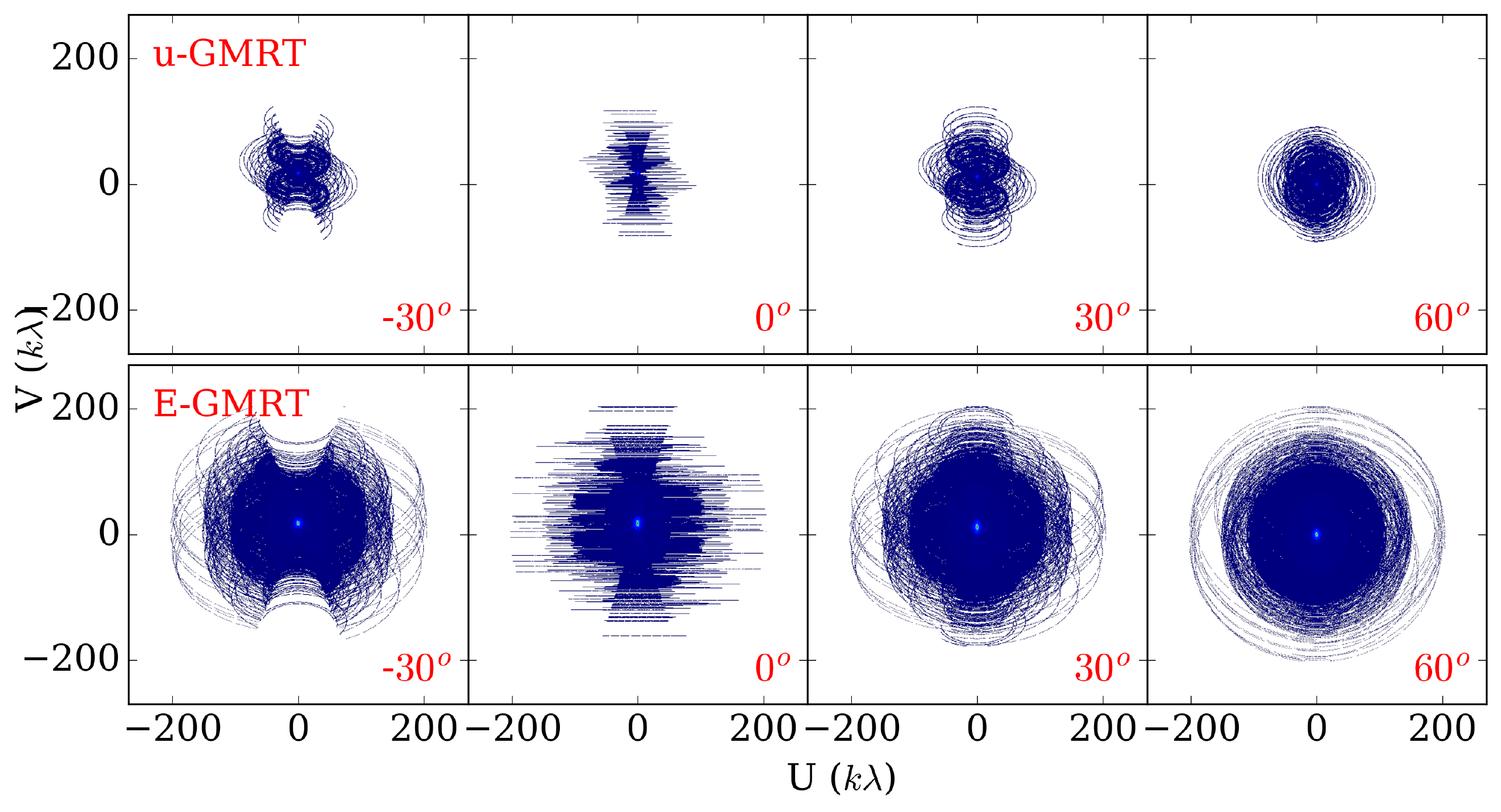}
\end{center}
\caption{A comparison between the single-channel U-V coverage of the full GMRT and the full EGMRT, i.e. 
with baselines out to 25~km for the GMRT and to 50~km for the EGMRT for the four declinations, 
for a full-synthesis observing run at 1.2~GHz.
\label{fig:cov-full5}}
\end{figure*}


\subsubsection{Optimization for FWHM~$= 5.0$~km, i.e. $b_{\rm max} = 15.0$~km}
\label{sec:bmax15km}

The next step, the optimization for FWHM~$=5.0$~km, included all the current GMRT antennas as fixed antennas, as each 
GMRT antenna yields a number of baselines of length $< 15$~km. We also included the 15 EGMRT antennas obtained from the 
earlier two optimizations as fixed antennas in the optimization. The optimization used the tomographic projection 
approach and did not include any land constraints.

Fig.~\ref{fig:rms3}[A] shows the Residual RMS plotted against the number of antennas for $\delta = -30^\circ$, $0^\circ$,
$+30^\circ$, and $+60^\circ$; the dashed vertical line is for 15 new antennas, above which the improvement in Residual
RMS with added antennas flattens. Fig.~\ref{fig:rms3}[B] shows the Residual RMS plotted against declination for the 
best configurations obtained with 15 new antennas for each of the four declinations. The best overall performance is 
obtained for the configuration with $\delta = 0^\circ$, shown in Fig.~\ref{fig:rms3}[C]. 
Fig.~\ref{fig:cov-full3} shows a comparison between the U-V coverage of the GMRT array with baselines out to 
$b_{\rm max} = 15.0$~km and the best EGMRT array (with 20 GMRT antennas and 30 new antennas), for 
the four declinations, and a full-synthesis run.

\subsubsection{Optimization for FWHM~$= 15.0$~km, i.e. $b_{\rm max} = 25.0$~km}
\label{sec:bmax25km}

Next, we fixed the locations of the 30 new antennas along with the 30 existing GMRT antennas, and carried out 
the optimization, using tomographic projection, for FWHM~$=15$~km, with $b_{\rm max} = 25.0$~km. Again, no 
land constraints were used in the optimization. Fig.~\ref{fig:rms4}[A] shows the derived Residual RMS plotted versus 
the number of added antennas for the four different declinations; the dashed vertical line is at 15 new antennas,
where the decline in the Residual RMS with new antennas slows down. Fig.~\ref{fig:rms4}[B] shows the Residual RMS 
plotted versus declination for the best configurations with 15 new antennas; we find that the configuration 
with $\delta = +60^\circ$, shown in Fig.~\ref{fig:rms4}[C] gives the best overall performance. Fig.~\ref{fig:cov-full4} compares the U-V coverage of the GMRT array with baselines out to $b_{\rm max} = 25.0$~km 
and the best above EGMRT array (with 30 GMRT antennas and 45 new antennas), for a full-synthesis 
observing run, for the different declinations. 

\subsubsection{Optimization for FWHM~$= 25.0$~km, i.e. $b_{\rm max} = 50.0$~km}
\label{sec:bmax50km}

Finally, we carried out the optimization for FWHM~$=25.0$~km, fixing the locations of the 30 GMRT antennas and the 
45 new EGMRT antennas, and again using the tomographic projection approach without land constraints, with $b_{\rm max} = 50$~km.
Fig.~\ref{fig:rms5}[A] shows the Residual RMS plotted versus the added number of antennas for the four different declinations;
the dashed vertical line is at 11 new antennas, beyond which the decline in Residual RMS with added antennas appears to 
flatten out. Fig.~\ref{fig:rms5}[B] shows the Residual RMS of the four best configurations obtained with 11 new antennas 
plotted versus declination; the configuration obtained from $\delta = +60^\circ$, shown in Fig.~\ref{fig:rms5}[C], yields 
the best overall performance. The U-V coverage of this array is shown in Fig.~\ref{fig:cov-full5} for a full-synthesis run, 
for the four different declinations. We note that the GMRT array only provides baselines out to $\approx 25$~km. With 
$b_{\rm max} \approx 50$~km, the EGMRT array would have a synthesized beam smaller by a factor of $\approx 2$, and would 
hence have a confusion noise lower by a factor of $\approx 10$ than that of the uGMRT at the same observing frequency.

\section{Discussion}
\label{sec:results}

\begin{figure*}
\begin{center}
	\includegraphics[height=3.0in]{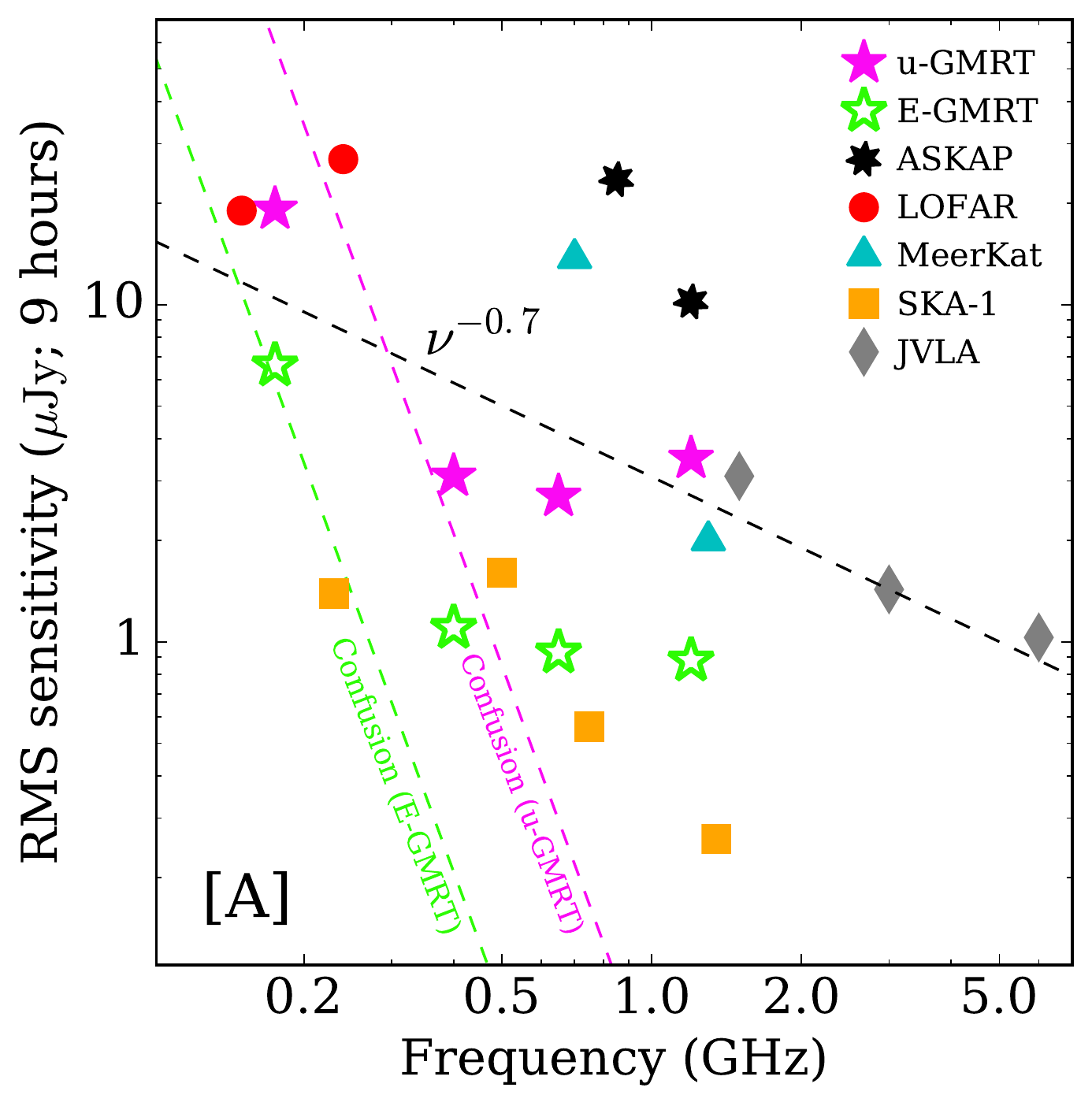}
	\includegraphics[height=3.0in]{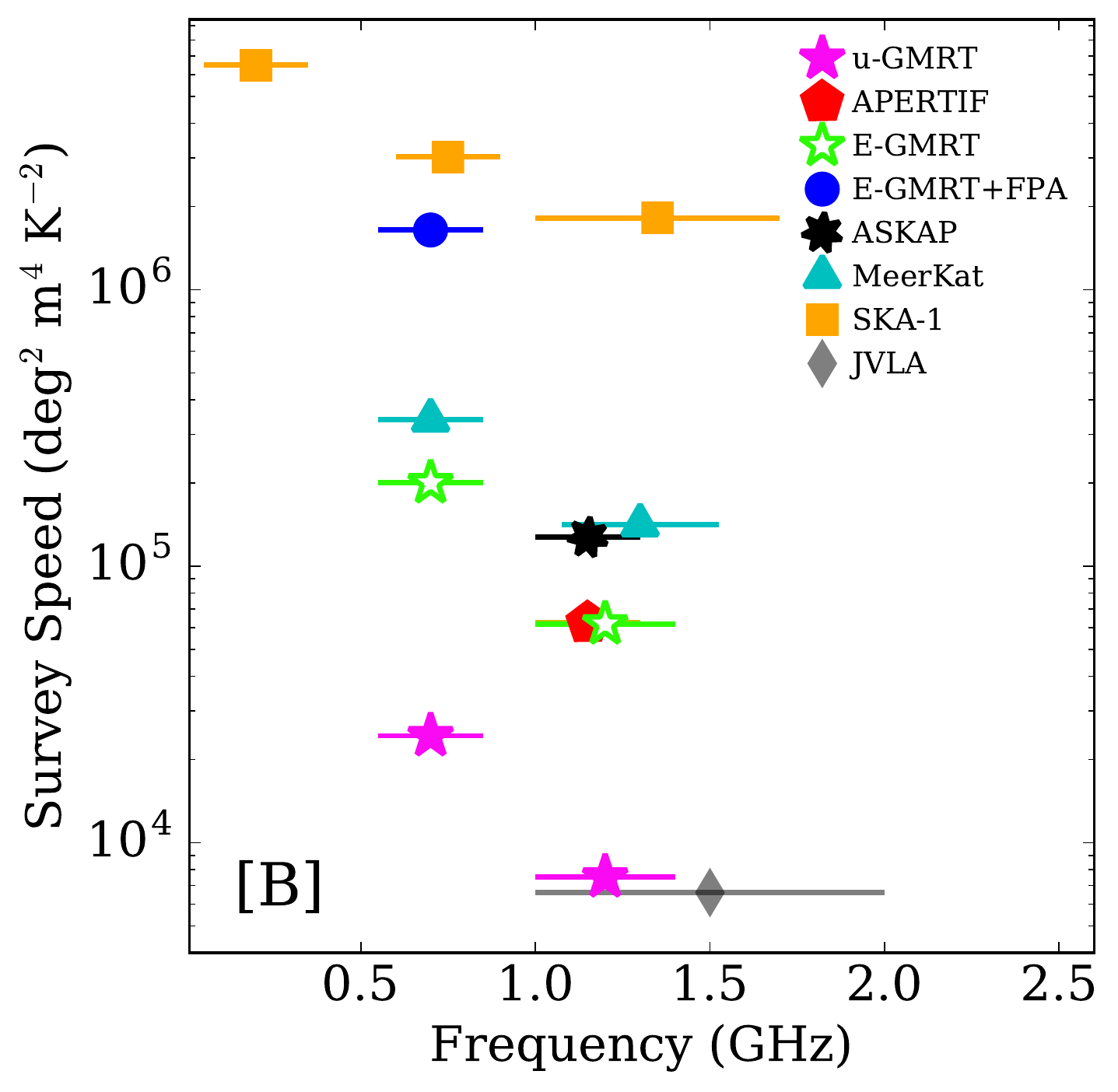}
\end{center}
\caption{[A]~Left panel: The $1\sigma$ continuum noise of the EGMRT compared with that of the other 
radio interferometers of Fig.~\ref{fig:ugmrt} (the uGMRT, the JVLA, LOFAR, MeerKAT, ASKAP, and the 
SKA-1) for a 9-hour full-synthesis integration. The green and magenta dashed lines show the $1\sigma$ 
confusion noise for, respectively, the EGMRT and the uGMRT, at the different observing 
frequencies. It is clear that source confusion will be a limiting factor for the EGMRT 
only in its lowest frequency band ($125-250$~MHz), where the sensitivity is likely to anyway be limited 
by systematic effects, rather than thermal noise. Note that the $1\sigma$ continuum noise values for 
ASKAP and MeerKAT include confusion noise, also extrapolated from \citet{condon12}.
[B]~Right panel: The survey speed figure of merit, in deg.$^2$~m$^4$~K$^{-2}$ \citep[e.g.][]{dewdney15}, 
of the EGMRT compared with that of other present or planned radio interferometers. For the 
EGMRT, we have considered two possibilities: the open green stars (``EGMRT'') refer to 
single-pixel feeds on all 86 antennas, while the solid blue circles (``EGMRT+FPA'') assume FPAs 
covering $550-850$~MHz installed on the 45~antennas within $\approx 2.5$~km of the central square.
	\label{fig:egmrt}}
\end{figure*}

We thus find that adding 56 new antennas to the GMRT array, 30 antennas at distances $\lesssim 2.5$~km
from the GMRT central square, and 26 antennas more distant from the central square (at distances $\approx 7.5-25$~km), 
would significantly improve the U-V coverage of the array. The resulting baselines would, for declinations 
$\delta \approx -30^\circ - +60^\circ$, yield a U-V coverage close to a 2-D Gaussian distribution with FWHM's 
of $\approx 0.5$~km, $\approx 1.7$~km, $\approx 5.0$~km, $\approx 15$~km, and $\approx 25$~km, depending 
on the maximum baseline used in the imaging process. The longitudes and latitudes of the new EGMRT 
antennas are listed in Table~\ref{table:egmrt} in the Appendix, where each pair of longitude and latitude columns 
refers to antenna locations obtained in optimizations for increasing  FWHM's ($0.5-25$~km) of the U-V coverage.

Fig.~\ref{fig:egmrt}[A] compares the continuum sensitivity of the proposed EGMRT array to that of the other 
radio interferometers of Fig.~\ref{fig:ugmrt}, with the EGMRT values shown as open green stars. We assume 
that the EGMRT will have 86 dishes of 45-m diameter, with baselines out to $\approx 50$~km, and that
the instantaneous correlated bandwidth and the receiver sensitivity will be the same as that of the uGMRT.
It is clear that the EGMRT point-source sensitivity would be similar to that of the SKA-1 over 
$\approx 0.3-1.4$~GHz, and would be significantly better than that of any other radio telescope over its 
entire operating frequency range.

Fig.~\ref{fig:egmrt}[B] compares the survey speed of the proposed EGMRT with present and 
proposed radio interferometers at frequencies $\lesssim 2$~GHz. For this comparison, we have 
used the survey speed figure of merit of \citet{dewdney15}, defined as $\rm (A_{eff}/T_{sys})^2\Omega$
(in deg.$^2$~m$^4$~K$^{-4}$), where $\rm A_{eff}$ is the effective area of the array, $\rm T_{sys}$ is its 
system temperature, and $\Omega$ is its instantaneous field-of-view. Two configurations are shown 
for the EGMRT, the first (``EGMRT+FPA'') with FPAs covering $550-850$~MHz installed on the 
45~antennas within 5~km baselines, and the second (``EGMRT'') with single-pixel feeds on all 86 
antennas. It is clear that the survey speed of the first EGMRT configuration, with FPAs at $550-850$~MHz 
installed on 45~antennas, would be better than that of all present radio interferometers, and 
comparable to that of the SKA-1 in the same frequency band.

\section{Summary}

We have discussed three possible expansions of the GMRT to improve its surface brightness sensitivity, field of view 
(and, hence, survey speed), and confusion limit, and to retain its status as the premier low-frequency ($\lesssim 1$~GHz)
radio interferometer beyond the next decade. The three expansions involve adding antennas at short baselines (for 
surface brightness sensitivity) or on long baselines (to reduce the confusion noise), or replacing the GMRT single-pixel
feeds with FPAs to significantly increase the field of view. The primary science drivers of the three expansions are 
very different: adding FPAs would enable searches for \hii\ emission from high-$z$ galaxies and wide-field pulsar 
and continuum surveys, the short-baseline expansion would significantly improve the GMRT's capabilities for the 
mapping of extended radio emission in the Galactic Plane, in galaxy clusters, etc., while the long-baseline expansion 
would lower the confusion limit, allowing deep extra-galactic continuum studies. These will be discussed in detail 
in future science papers. To retain the multi-science capabilities of the GMRT, we propose that it would be best 
to not focus on a single expansion strategy, but to add antennas on both short and long baselines, and to also 
add FPAs on a subset of the antennas, on baselines $\lesssim 2.5$~km from the GMRT central square. Such a strategy 
would retain scientific flexibility, while significantly improving the capabilities of the expanded array for all 
the above science goals.

To achieve the above goal, we then identified the optimal locations for the new antennas of the proposed array,
following an inside-out approach aimed at obtaining a well-behaved synthesized beam over a wide range of angular 
scales. While the final antenna locations of the EGMRT, especially on the long baselines, will depend on 
practical constraints, including land availability, optical fibre connectivity, roads, etc., the present configuration
provides a benchmark array, relative to which one can evaluate the final EGMRT configuration. The array optimization
was carried out by choosing antenna locations so as to obtain a U-V coverage as close as possible to a 2-D circular 
Gaussian distribution (which would yield a 2-D Gaussian synthesized beam), with different FWHM's (0.5~km, 1.7~km, 
5~km, 15~km, and 25~km), and for a wide range of declinations ($-30^\circ$ to $+60^\circ$). The optimization of 
antenna locations was carried out using two strategies, random-sampling for antennas located close to the GMRT 
central square, i.e. for FWHM's~$\leq 1.7$~km, and tomographic projection for FWHM's~$\geq 5$~km. We find that 
the requirement that the U-V coverage be close to a 2-D circular Gaussian can be met by adding $8$ antennas within 
the central $\approx 0.5$~km region of the GMRT central square, $7$ antennas at distances $\approx 0.5-1$~km from 
the central square, $15$~antennas $\approx 1-2.5$~km from the central square, $15$~antennas $\approx 7.5$~km from the 
central square and $11$~antennas $\approx 25$~km from the central square. The 30 new EGMRT antennas within 
$\approx 2.5$~km of the central square would contribute significantly towards improving the surface brightness 
sensitivity, as well as the sensitivity for pulsar surveys and searches for redshifted \hii\ emission, while the 
11~antennas on baselines $\gtrsim 25$~km would contribute to improving the confusion limit for extragalactic 
continuum studies. We further propose to install FPAs on the new EGMRT and existing GMRT antennas within a 
$\approx 5$~km region around the GMRT central square, to achieve a large field of view for this group of 45
antennas, significantly improving the survey speed of the array.

Finally, we compared the sensitivity and survey speed of the proposed new EGMRT array, with 56 new antennas, 
to that of existing and planned radio interferometers. We find that the point-source sensitivity of the EGMRT 
would be similar to that of the proposed SKA-1 at frequencies $300-1000$~MHz, and significantly better than 
that of all other existing and planned interferometers at frequencies $\lesssim 1.4$~GHz. Similarly, the survey 
speed of the EGMRT sub-array equipped with FPAs would be comparable to that of the SKA-1 at a similar frequency,
and higher than that of any other existing or planned radio interferometer. The proposed expansions would thus 
allow the GMRT to retain its status as the premier low-frequency radio interferometer in the world well beyond 
the next decade, and into the SKA era.

\bibliographystyle{mn2e}
\bibliography{ms_new}

\appendix
\vskip -10in
\section{}
\begin{table*}
\caption{The coordinates of the new EGMRT antennas}
  \begin{tabular}{|c|c|c|c|c|c|c|c|c|c|}
    \hline
    \multicolumn{2}{|c|}{0.5 km} &
    \multicolumn{2}{|c|}{1.7 km} &
    \multicolumn{2}{|c|}{5 km} &
    \multicolumn{2}{|c|}{15 km} &
    \multicolumn{2}{|c|}{50 km} \\
Long. ($^\circ$) & Lat. ($^\circ$) & Long. ($^\circ$) & Lat. ($^\circ$) & Long. ($^\circ$) & Lat. ($^\circ$) & Long. ($^\circ$) & Lat. ($^\circ$) & Long. ($^\circ$) & Lat. ($^\circ$) \\
    \midrule
74.0507725 & 19.0900447 & 74.0463238 & 19.0928289 & 74.0664129 & 19.0999286 & 73.9609521 & 19.1032398 & 74.0664124 & 18.8880061 \\
74.0509136 & 19.0894700 & 74.0429033 & 19.0825301 & 74.0673838 & 19.0928796 & 74.1478274 & 18.9959265 & 74.2835108 & 19.1559832 \\
74.0497636 & 19.0898942 & 74.0486044 & 19.0987914 & 74.0366723 & 19.1047456 & 73.9672950 & 19.0201074 & 74.1844691 & 18.8994765 \\
74.0482780 & 19.0896067 & 74.0486044 & 19.1004175 & 74.0348558 & 19.1015569 & 74.0227518 & 19.2076507 & 74.2757907 & 19.0063713 \\
74.0505633 & 19.0886720 & 74.0503149 & 19.0933710 & 74.0345159 & 19.0949776 & 74.1580258 & 19.0830605 & 74.0590555 & 19.2845531 \\
74.0513364 & 19.0901220 & 74.0503149 & 19.0857825 & 74.0546174 & 19.1089636 & 73.9727482 & 18.9983853 & 73.8157773 & 19.1618540 \\
74.0491589 & 19.0908299 & 74.0497447 & 19.0982494 & 74.0508886 & 19.1093734 & 74.1629440 & 19.1133690 & 73.8299016 & 18.9894072 \\
74.0521847 & 19.0900995 &            &            & 74.0585228 & 19.1077731 & 74.1620281 & 19.0395763 & 74.2884355 & 19.1360591 \\
           &            &            &            & 74.0673717 & 19.0918209 & 74.1147037 & 19.2212894 & 73.8146510 & 19.0262421 \\
           &            &            &            & 74.0450750 & 19.1089053 & 74.1235203 & 18.9773132 & 74.1148953 & 18.8873737 \\
           &            &            &            & 74.0669867 & 19.0973430 & 74.1532751 & 19.0351790 & 73.8052941 & 19.1077472 \\
           &            &            &            & 74.0659674 & 19.1013593 & 74.1588607 & 19.0598107 &            &             \\
           &            &            &            & 74.0359408 & 19.0925731 & 74.1496416 & 19.1847213 &            &             \\
           &            &            &            & 74.0566546 & 19.0797704 & 74.1642385 & 19.1511805 &            &             \\
           &            &            &            & 74.0632511 & 19.1052599 & 73.9635819 & 19.0694059 &            &             \\    
  \hline
  \end{tabular}
  \label{table:egmrt}
\end{table*}

\end{document}